\documentclass[fleqn,usenatbib]{mnras}

\usepackage{textcomp}
\usepackage{amsmath}

\usepackage[T1]{fontenc}

\DeclareRobustCommand{\VAN}[3]{#2}
\let\VANthebibliography\thebibliography
\def\thebibliography{\DeclareRobustCommand{\VAN}[3]{##3}\VANthebibliography}

\usepackage{graphicx}	
\usepackage{amsmath}	
\usepackage{amssymb}	
\usepackage{xcolor}
\usepackage{nicefrac}

\newcommand{\be}{\begin{equation}}
\newcommand{\ee}{\end{equation}}

\newcommand{\ba}{\begin{eqnarray}}
\newcommand{\ea}{\end{eqnarray}}
\newcommand{\NS}{neutron star}
\newcommand{\psr}{PSR~B$1259-63$}
\newcommand{\frbs}{FRB~$180916.\mathrm{J}0158+65$}
\newcommand{\frbl}{FRB~$121102$}

\DeclareSymbolFont{matha}{OML}{txmi}{m}{it}
\DeclareMathSymbol{\varv}{\mathord}{matha}{118}

\title[Periodic FRB in eccentric binaries]{Formation of periodic FRB in binary systems with eccentricity}

\author[M.V.~Barkov \& S.B.~Popov]{
Maxim V. Barkov,$^{1}$\thanks{E-mail: barkov@inasan.ru}
Sergei B. Popov$^{2}$
\\
$^{1}$ Institute of Astronomy, Russian Academy of Sciences, Moscow, 119017 Russia \\
$^2$ Sternberg Astronomical Institute, Lomonosov Moscow State University, 119234 Russia
}

\date{Accepted XXX. Received YYY; in original form ZZZ}

\pubyear{2022}

\begin{document}
\label{firstpage}
\pagerange{\pageref{firstpage}--\pageref{lastpage}}
\maketitle

\begin{abstract}
 Long-term periodicity in the rate of flares is observed for two repeating sources of fast radio bursts (FRBs). 
 In this paper we present a hydrodynamical modeling of a massive binary  consisting of a magnetar and an early-type star.  We model the interaction of the pulsar wind from the magnetar  with an intense stellar wind.   It is shown that only during a fraction of the orbital period radio emission can escape the system. This explains the duty cycle of the two repeating FRB sources with periodic activity. The width of the transparency window depends on the eccentricity, stellar wind properties, and the viewing angle.  To describe properties of the known sources it is necessary to assume large eccentricities $\gtrsim 0.5$. We apply the maser cyclotron mechanism of the radio emission generation to model spectral properties of the sources. The produced spectrum is not wide: $\Delta \nu/\nu \sim 0.2$ and the typical frequency depends on the radius of the shock where the emission is generated. The shock radius changes along the orbit. This, together with changing parameters of the medium, allows us to explain the frequency drift during the phase of visibility.
 Frequency dependence of the degree of polarization at few GHz can be a consequence of a small scale turbulence in the shocked stellar wind.
 It is much more difficult to explain huge ($\sim 10^5$~[rad/m$^2$]) and variable value of the rotation measure observed for FRB 121102. 
  We suggest that this can be explained if the
supernova explosion which produced the magnetar happened near a dense interstellar cloud with $n \sim100$ cm$^{-3}$.

\end{abstract}




\section{Introduction}
\label{sec:intro}

Since announcement of the first fast radio burst (FRB) \citep{2007Sci...318..777L}, and especially, after the paper by \cite{2013Sci...341...53T}, numerous hypothesis about the nature of these transients were proposed (see a list in \citealt{2019PhR...821....1P}). Step by step, new observational results helped to eliminate some of those models, on other hand providing support to others (see reviews on FRBs in
\citealt{2020Natur.587...45Z, 2021Univ....7...76N, 2022arXiv220314198X}).

Several models were rejected on the basis of the rate of FRBs. Already first papers \citep{2007Sci...318..777L, 2013Sci...341...53T} on the subject suggested that the rate is above $\sim 10^3$ per sky per day. Later observations agreed with this estimate, and presently, having several hundreds detected events, not counting multiple flares of repeating sources (see data in the on-line catalogue http://wis-tns.org, \citealt{2016PASA...33...45P}), the rate is estimated to be $\sim 10^4$ per day for the limiting flux above $\sim 0.1$Jy (see  \citealt{2017AJ....154..117L, 2019A&ARv..27....4P}). This obviously excludes many scenarios of the origin of FRBs (see \citealt{2020MNRAS.494..665L} and references therein), for example,  coalescence of neutron stars (NSs) and black holes (BHs) (white dwarf-NS coalescence, which can be quite numerous, are also excluded, see \citealt{2019JHEAp..24....1K}). 

Early discussion, when the situation with the origin of FRBs was absolutely unclear, was mainly related to types of sources: exotics (cosmic strings, evaporating black and white holes, etc.), known transients (supernovae, GRBs, coalescence, etc.), known types of neutron star activity (magnetars, giant pulses), exotics related to NSs (deconfinement, falling asteroids, collapse to a BH). At this situation, it was pre-mature to discuss details of possible emission mechanism. However, with appearance of more and more detailed observations such studies became desirable. 

In 2014 the first FRB was detected and identified in real time \citep{2015MNRAS.447..246P}. This allowed to trigger multiwavelength observations, but no counterparts were detected. This excluded models involving supernovae (SNae) and $\gamma$-ray bursts (GRBs), as well as some other bright transients.  

Major progress was related to discovery of repeating sources of FRBs. Repeated bursts of the most prolific source -- \frbl{} \citep{2014ApJ...790..101S} -- were detected initially thanks to large collecting area of the Arecibo antenna \citep{2016Natur.531..202S}. This result not only excluded all catastrophic scenarios, but also allowed -- for the very first time, -- to obtain precise localization of the source (via interferometric observations), to identify the host galaxy, and thus, to prove the extragalactic nature of this phenomenon, providing exact energetics of FRBs \citep{2016Natur.530..453K,  2017Natur.541...58C, 2017ApJ...834L...7T, 2017ApJ...843L...8B}. 

In the following years $>20$ repeating sources were identified (see, e.g. \citealt{2020ApJ...891L...6F} and references therein). Some of them demonstrated just few flares, on other hand, several produced many bursts. Frequent activity allowed programs of multiwavelength monitoring. Despite many efforts, no counterparts were detected at any energy range, up to TeVs \citep{2017ApJ...846...80S}. The same can be said about non-repeating sources. Programs of observations with $\gamma$-ray monitors \citep{2016MNRAS.460.2875Y, 2017ApJ...842L...8X}, wide field optical telescopes \citep{2017MNRAS.472.2800H}, as well as X-ray telescopes, and LOFAR \citep{2014A&A...570A..60C, 2016MNRAS.459.3161C} produced zero result: no FRBs were detected at other wavelengths.

Localization of FRBs is possible not only for repeating sources. At the moment $\sim 20$ sources are localized and this number is dominated by single events, mainly thanks to ASKAP \citep{2020ApJ...895L..37B}. Host galaxies of FRBs show variety of properties: some have strong starformation rate, some --- not (see the on-line catalogue of FRB hosts at http://frbhosts.org \citealt{2020ApJ...903..152H}). This information also allows to make critical analysis of different scenarios of the origin of FRBs (see \citealt{2020ApJ...899L...6L} and references therein). 

\begin{figure}
	\includegraphics[width=\columnwidth]{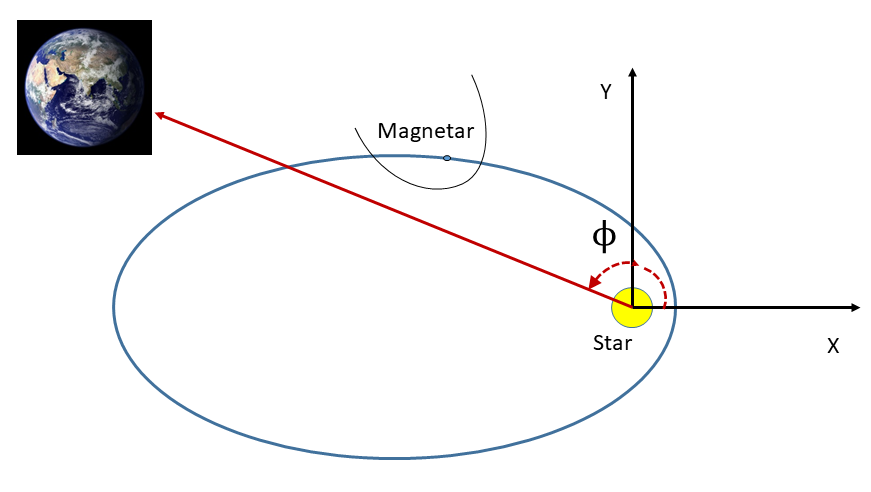}
    \caption{Sketch of a binary system with a magnetar producing fast radio bursts. {The magnetar is surrounded by a bow shock formed due to winds collision. The normal star is much heavier than the NS, thus the center of mass of the system roughly coincides with the position of the normal star.}  }
    \label{fig:sketct}
\end{figure}

\begin{figure*}
	\includegraphics[width=\columnwidth]{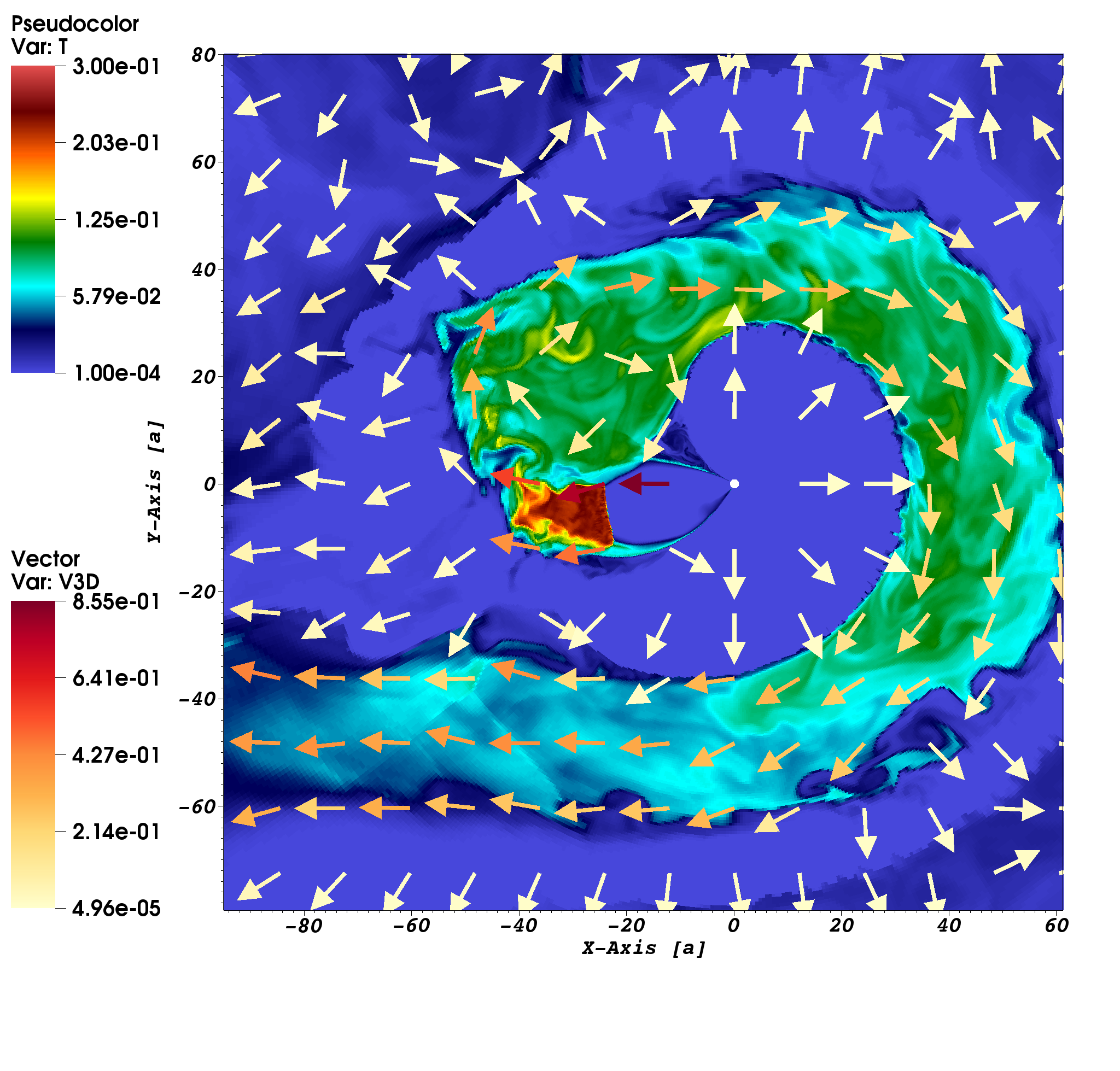}
	\includegraphics[width=\columnwidth]{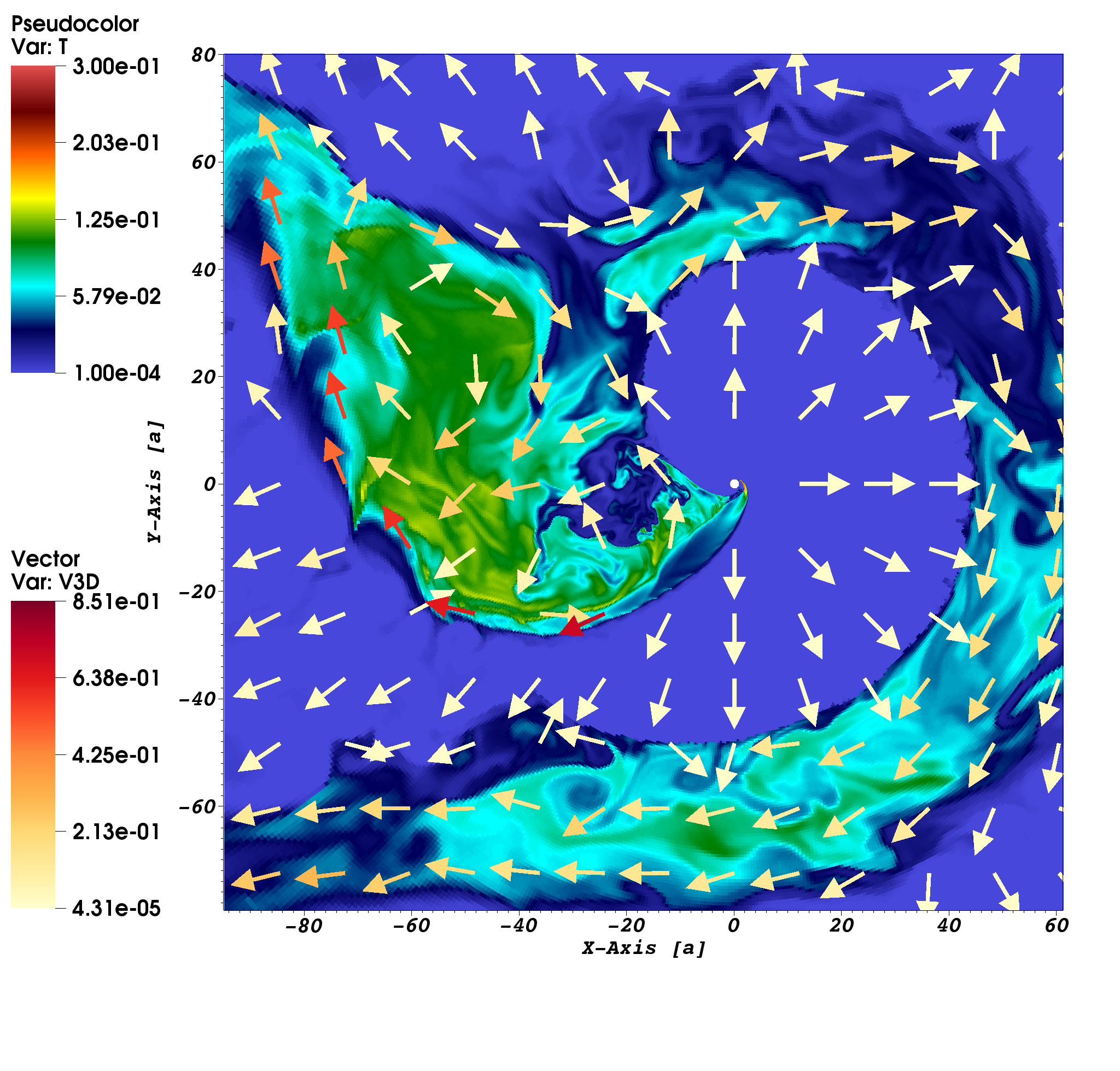}
    \caption{The dimensionless temperature (color) and velocity field (arrows) show two orbital phases near appoastron (right panel) and periastron (left panel). The appoastron is placed on the left side. The orbital period is 16 days and $\eta =0.1$.}
    \label{fig:Tdist}
\end{figure*}

\begin{figure*}
	\includegraphics[width=\columnwidth]{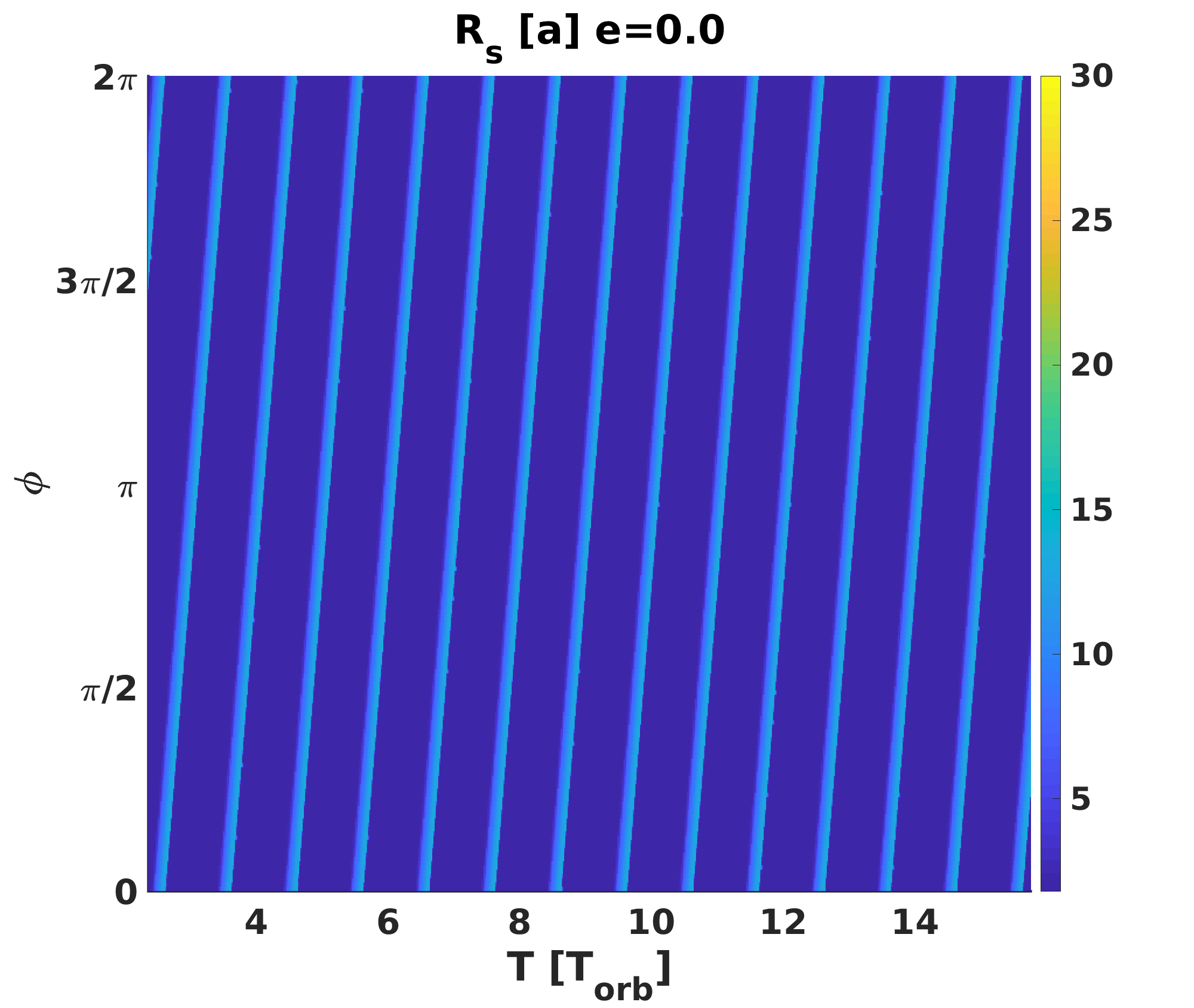}
	\includegraphics[width=\columnwidth]{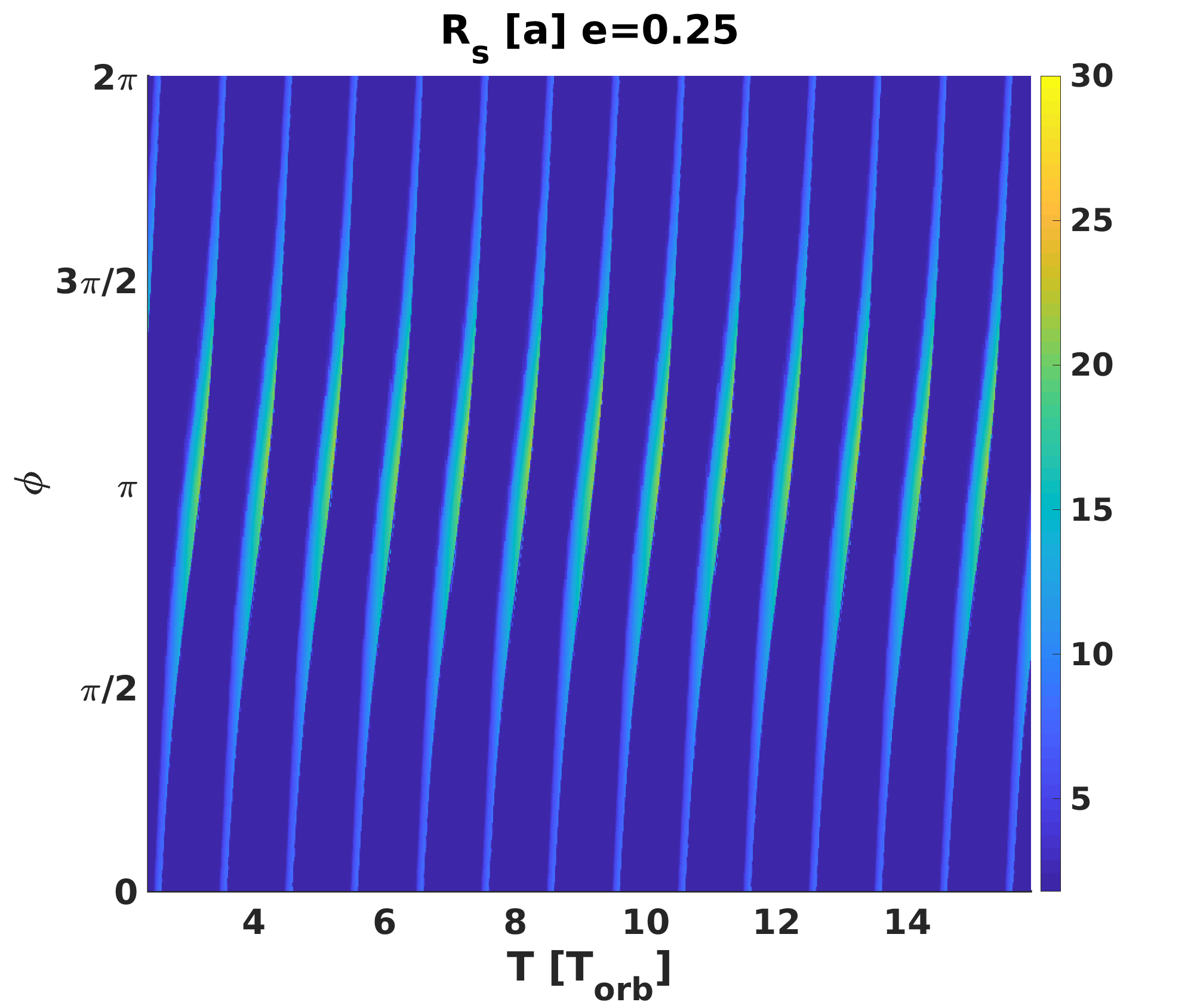}
	\includegraphics[width=\columnwidth]{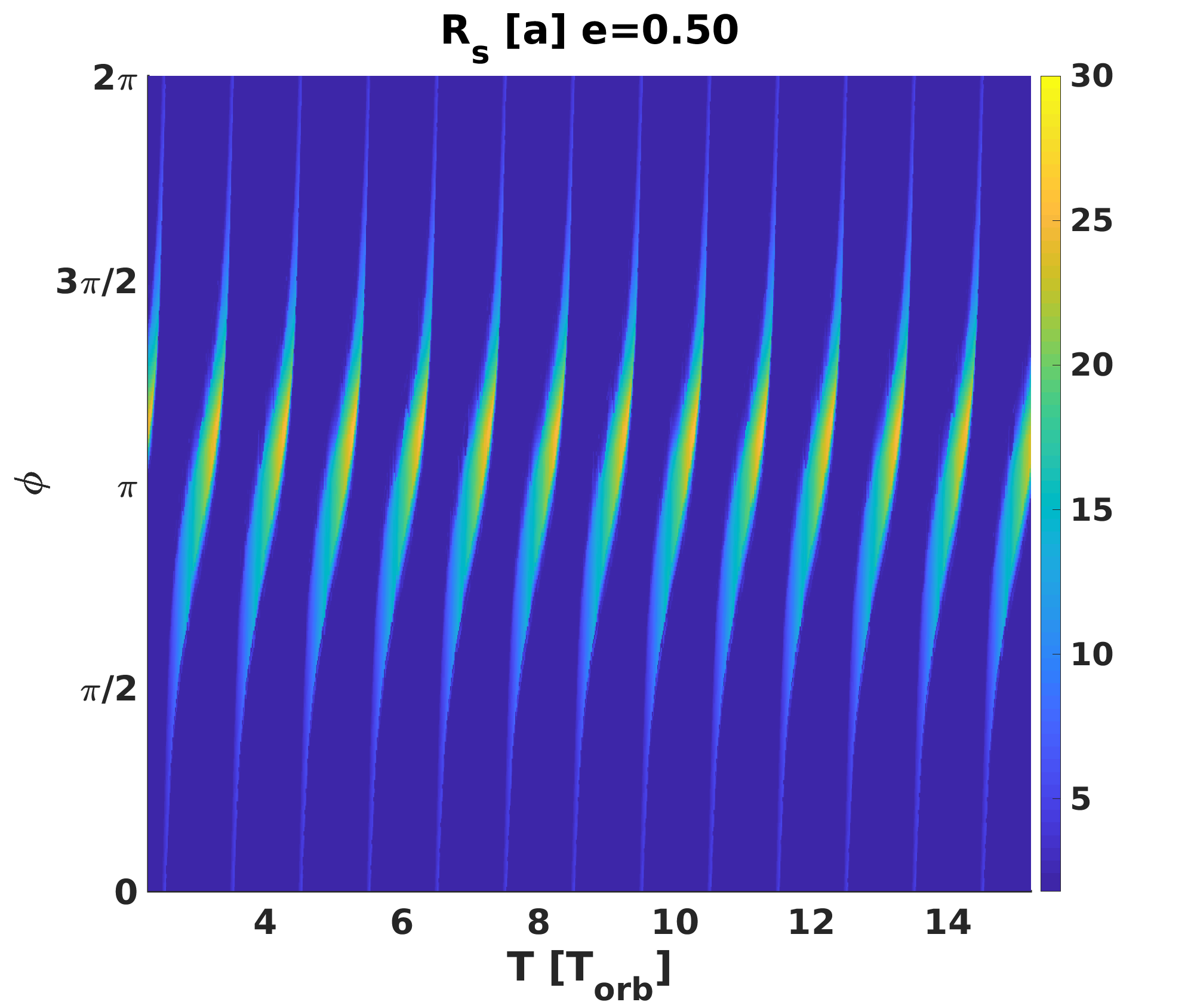}
	\includegraphics[width=\columnwidth]{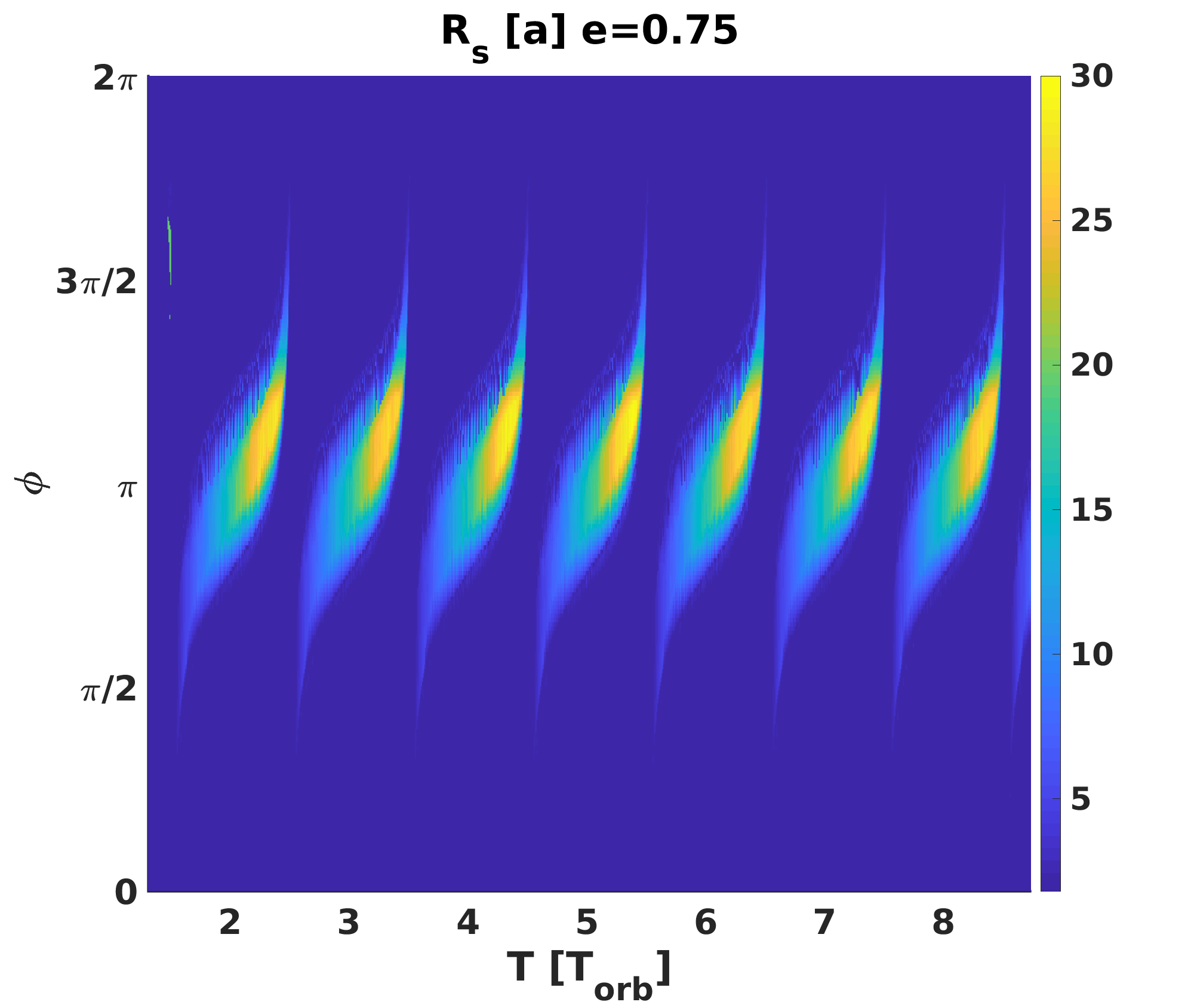}
    \caption{Dependence of the shock wave radius on the viewing angle $\phi$ and orbital phase for several orbital periods.
    The shock wave radius is normalized to the semi-major axis value. Four panels correspond to different eccentricities: $e=0.0$ (top-left panel), $e=0.25$ (top-right panel), $e=0.50$ (bottom-left panel), $e=0.75$ (bottom-right panel). 
    }
    \label{fig:RTphidist}
\end{figure*}

 In 2014 the first circular polarization measurements were made for FRB 140514 \citep{2015MNRAS.447..246P} and later results on  FRB110523 for which linear polarization were published \citep{2015Natur.528..523M}. 
 Now for $\gtrsim 20$ of sources, including several repeaters, polarization is measured (linear is observed more often than circular), see Table 2 in  \cite{2021Univ....7..453C}. 
 In several cases the degree of linear polarization is compatible with 100\% within uncertainty. 
 Rotation measure (RM) is measured in $\sim20$ cases, also see \cite{2021Univ....7..453C}.
 
 For quite a long time no detection at low frequencies has been known. Then, at first CHIME detected sources below 600 MHz \citep{2019Natur.566..230C}. Slightly later, thanks to high sensitivity of GBT (Green Bank Telescope) and SRT (Sardinia Radio Telescope)  FRBs were detected even below 400 MHz \citep{2020ApJ...896L..41C, 2020ApJ...896L..40P}. 
 Finally, LOFAR detected bursts of the repeating source FRB 180916 at $\gtrsim 110$~MHz \citep{2021ApJ...911L...3P}. A  recent review of low-frequency observations of FRBs can be found in \cite{2021Univ....8....9P}. However, bursts have never been detected simultaneously at significantly different frequencies (see discussion in \citealt{2021Univ....7..453C}).
 
 After few years of studies, two mainstream scenarios appeared. Both are related to NSs: magnetars and pulsars. Both scenarios include many various models of burst generation. General difference is that in the first approach the magnetic energy is emitted. In the second --- the rotational energy of a NS. Detailed studies in the framework of the magnetar scenario were initiated with the paper by \cite{2014MNRAS.442L...9L}. Later on, many authors contributed in this field (see \citealt{2016MNRAS.461.1498M, 2017MNRAS.468.2726K, 2017ApJ...843L..26B, 2020ApJ...896..142B, 2017ApJ...841...14M, 2019MNRAS.485.4091M,  2020ApJ...889..135L} and references therein).
 The pulsar scenario was also studied in several papers: \cite{2014PhRvD..89j3009K, 2015ApJ...807..179P, 2016MNRAS.457..232C, 2016MNRAS.458L..19C, 2016MNRAS.462..941L, 2016ApJ...824L..32P, 2018MNRAS.478.4348I}.
 Generally, in the first scenario radio bursts are related to magnetar giant flares, and in the second one analogues of the Crab giant pulses are discussed\footnote{Both possibilities were proposed immediately after the announcement of the first event \citep{2007arXiv0710.2006P}.}.
 
 Later studies demonstrated \citep{2017ApJ...838L..13L, 2019arXiv190103260L} that the pulsar scenario has significant drawbacks. And recently the main focus is on the magnetar scenario. In 2020, the latter framework got sudden support from observations of a Galactic source.
 
 For several years, one of the main arguments against the magnetar scenario was based on non-detection of radio bursts from Galactic magnetars, especially after the Dec. 2004 hyperflare \citep{2016ApJ...827...59T}.
 Galactic magnetars are known to be radio active: five sources demonstrated radio pulses with spin period or/and persistent radio emission \citep{2015RPPh...78k6901T}. However, nothing like FRBs were detected till 2019. 
 In 2019 millisecond duration radio bursts were detected from the Galactic magnetar XTE 1810-197 \citep{2019ApJ...882L...9M}. Still, these events do not look like twins of FRBs and can represent a different type of activity (see recent detailed study with simultaneous radio and X-ray observations in \citealt{2020arXiv200508410P}). 
 Situation changed in late April 2020, when a strong millisecond radio burst was detected from SGR 1935+2154 \citep{2020Natur.587...54C, 2020Natur.587...59B} coincident with a strong high-energy burst \citep{2020ApJ...898L..29M,  2021NatAs...5..378L, 2021NatAs...5..372R, 2021NatAs...5..401T}. 
 This discovery was accepted by many as the final confirmation that FRBs are related to magnetar flares \citep{2020MNRAS.498.1397L}.\footnote{Note, that also several studies demonstrated that statistical properties of some subsamples of FRB flares are similar to those of magnetars \citep{2020MNRAS.491.1498C, 2017JCAP...03..023W, 2019ApJ...879....4W}.} May be, such conclusion is pre-mature, but indeed, argumentation looks strong, and this gives a new boost to detailed analysis of different emission models in the framework of the magnetar scenario. 
 
 Now most of researches agree that FRBs are related to NSs, and majority is inclined in favor of magnetic energy source. Of course, the final proof might be related to identification of a high-energy counterpart for an extragalactic FRB corresponding to a giant magnetar flare. However, this is not that probable, as typical distances to FRBs are at least about few hundred Mpc, and with existing X/$\gamma$-ray monitors magnetar flares can be detected only closer than 100 Mpc \citep{2005MNRAS.362L...8L, 2006MNRAS.365..885P}. However, recent discovery of an FRB in a near-by galaxy M81 \citep{2021ApJ...910L..18B} gives some hope for future multi-wavelength identification.
 
 More clews can be found due to any kind of periodicity in the FRB emission. Since early years, one of the most tempting goal is detection of the periodicity related to the spin of compact object. Such feature can be found within burst structure, if periods are about few milliseconds, or it can be related to repetitions of bursts on timescale of seconds, as it was in the case of rotating radio transients (RRATs) \citep{2006Natur.439..817M}. Unfortunately, up to now this type of periodicity in FRBs is not known. However, some uncertain candidates exist \citep{2021arXiv210708463T}. 
 
 Nevertheless, analysis of repeating bursts allowed to identify some regularity in appearance of radio transients. CHIME observations identified 16.35 day period in activity of \frbs{} \citep{2020Natur.582..351C}. Activity is limited to a four-day period, the rest of time the source does not produce bursts.  Dispersion measure (DM) remains constant in all observations of the source. Immediately,  several papers with theoretical models of this periodicity appeared. The main ideas relate the period either to orbital motion \citep{2020ApJ...893L..39L, 2020ApJ...893L..26I,2021ApJ...920...54W}, or to precession \citep{2020ApJ...895L..30L, 2020ApJ...892L..15Z}. The idea that the observed periodicity is due a super-long spin period of a NS also has been proposed \citep{2020MNRAS.496.3390B}.  Periodicity with an order of magnitude longer time scale $\sim 160$ days was also reported for \frbl{} \citep{2020MNRAS.495.3551R}. 
 
 It was discussed that central sources in active repeating FRBs can have a different origin from the majority of non-repeating sources\footnote{Or, better say, sources with probably rarely repeating bursts, as long repetition periods are still possible.} (see, e.g. \citealt{2020ApJ...899L...6L} and references therein). 
 Note, that if the periodicity detected in two most prolific repeaters is due to orbital modulation, then it puts important constraints on the origin of sources \citep{2020RNAAS...4...98P}. In particular, it nearly excludes coalescence of two neutron stars 
 because it is quite improbable to form a triple system which can survive through two SN explosions and save the outer companion in an orbit with just a $\sim 16$-day period. Scenarios with accretion-induced collapse (AIC) of a white dwarf (WD) require a relatively low-mass normal outer companion (as it might outlive the WD progenitor) which cannot produce strong wind, compact objects as companions are also excluded for the same reason (no dense outflows). Thus, if intervals without detected bursts are due to absorption of a radio emission in a surrounding medium, then AIC is not a viable option. The same can be said about WD-NS coalescence. Then, the most probable conclusion might be that (presumable) magnetars in \frbl{} and \frbs{} were formed in a (more or less) normal core collapse supernova (SN). 
 
 In this work we develop a model in which activity of an FRB source is related to a magnetar, and periodicity is explained by the orbital motion on an eccentric orbit in a high mass binary. In some respect, the present study is an 
 extension  of the line of analysis initiated by \cite{2020ApJ...893L..39L}. In that paper the authors discussed the emission mechanism operating within the magnetar magnetosphere. An absorption mechanism  can be in operation if the source is situated inside the magnetar magnetosphere or at the distance of the shocked pulsar wind. All parameters, when possible, are taken to fit properties of the FRB 180916.J0158+65.

In the following section (Sec.~\ref{sec:mod}) we describe the model used in this study. Results are presented in Sec.~\ref{sec:results}.  In Sec.~\ref{sec:Emission} we model and analyse a periodic formation of a transparent zone near the apoastron  of the orbital motion. In this region the pulsar's `tail' extends up to tens of the semi-major axes of the system. 
 In the framework of such  model, the emission mechanism is based on the synchrotron maser effect suggested for FRBs by \cite{2014MNRAS.442L...9L}. \cite{2022ApJ...927....2K} suggested and justified  that the necessary conditions for an  FRB pulse formation can be reached in such circumstances. 
The final Sec.~\ref{sec:dis} contains a brief discussion and conclusions.

\begin{figure*}
	\includegraphics[width=\columnwidth]{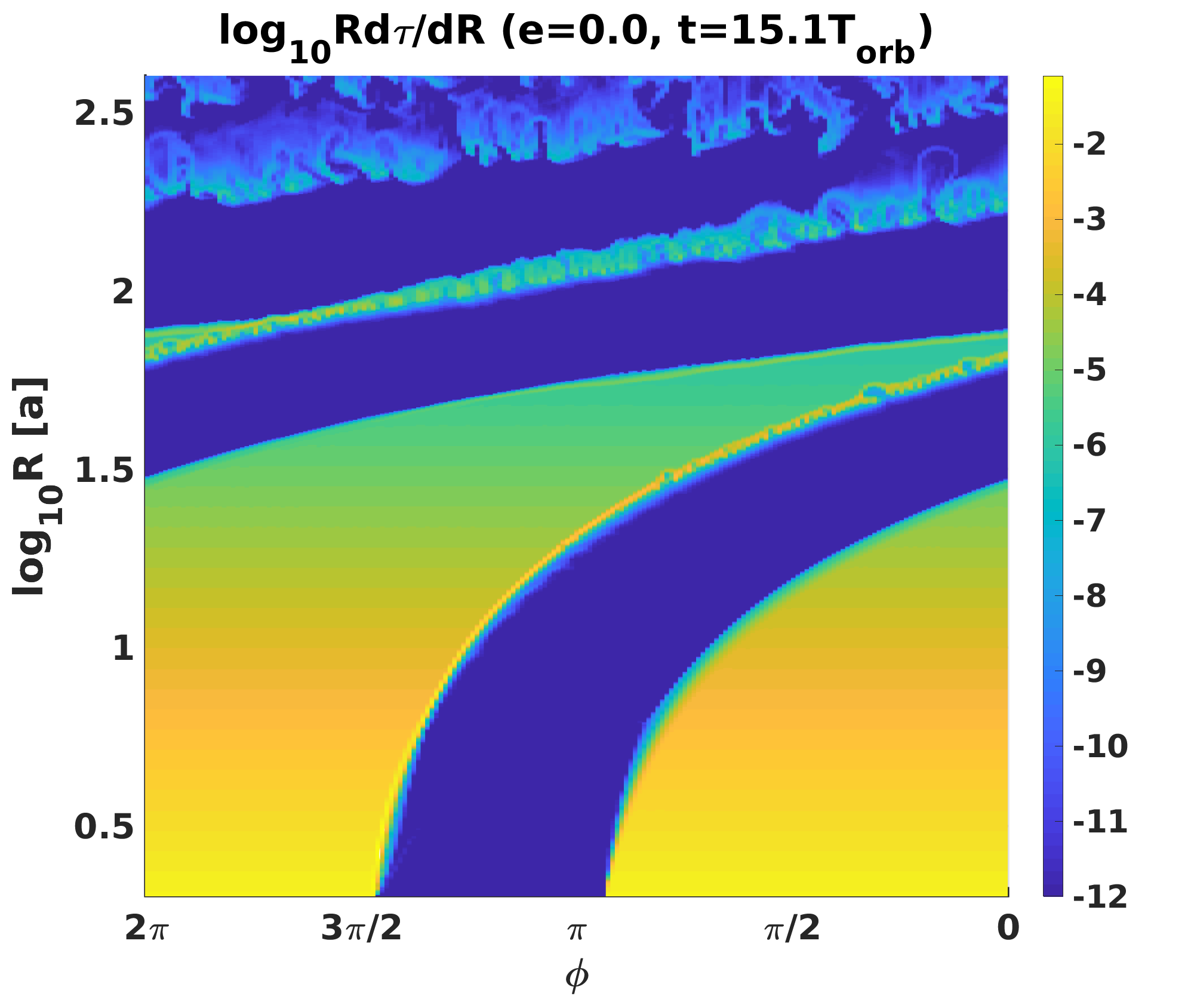}
	\includegraphics[width=\columnwidth]{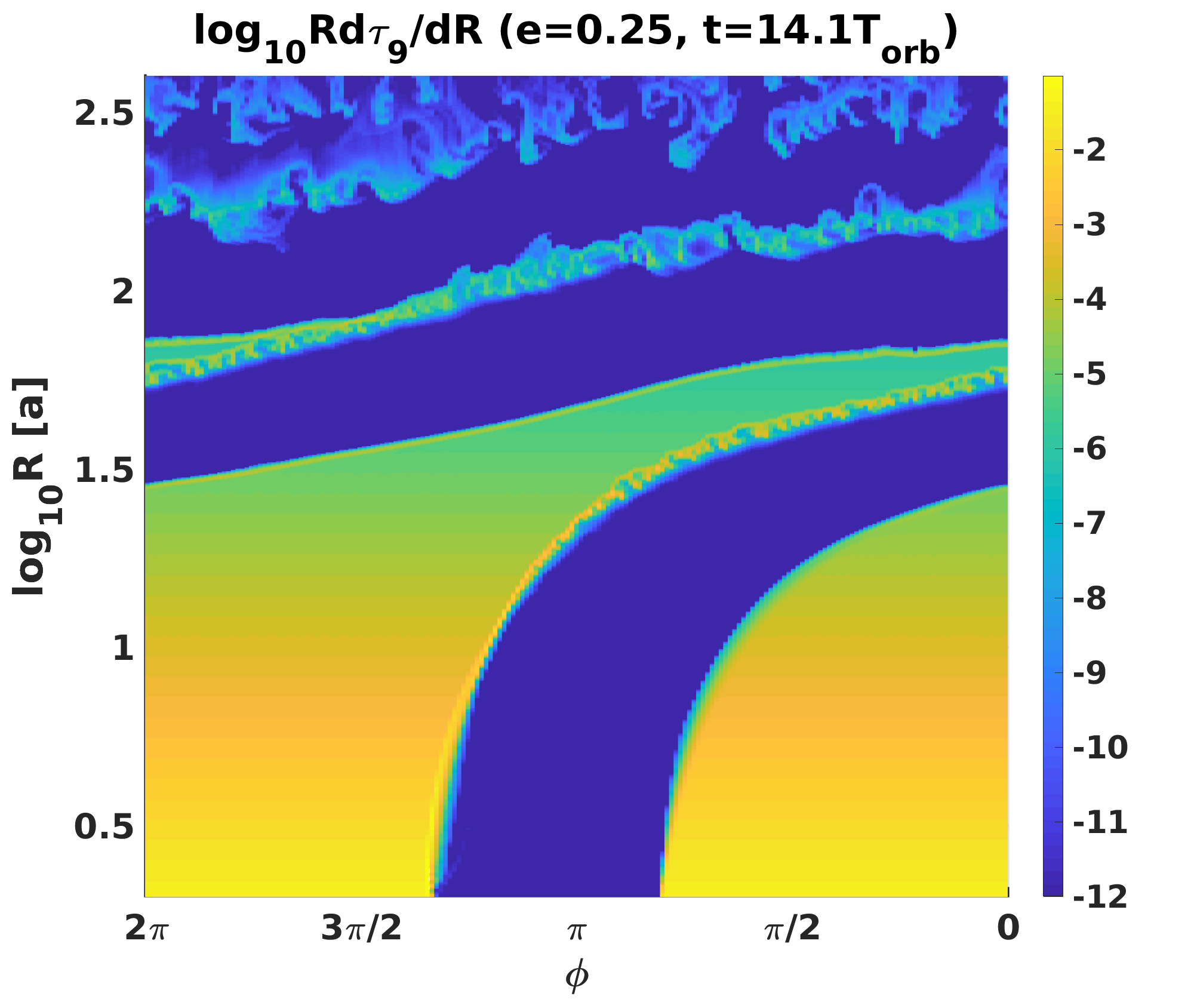}
	\includegraphics[width=\columnwidth]{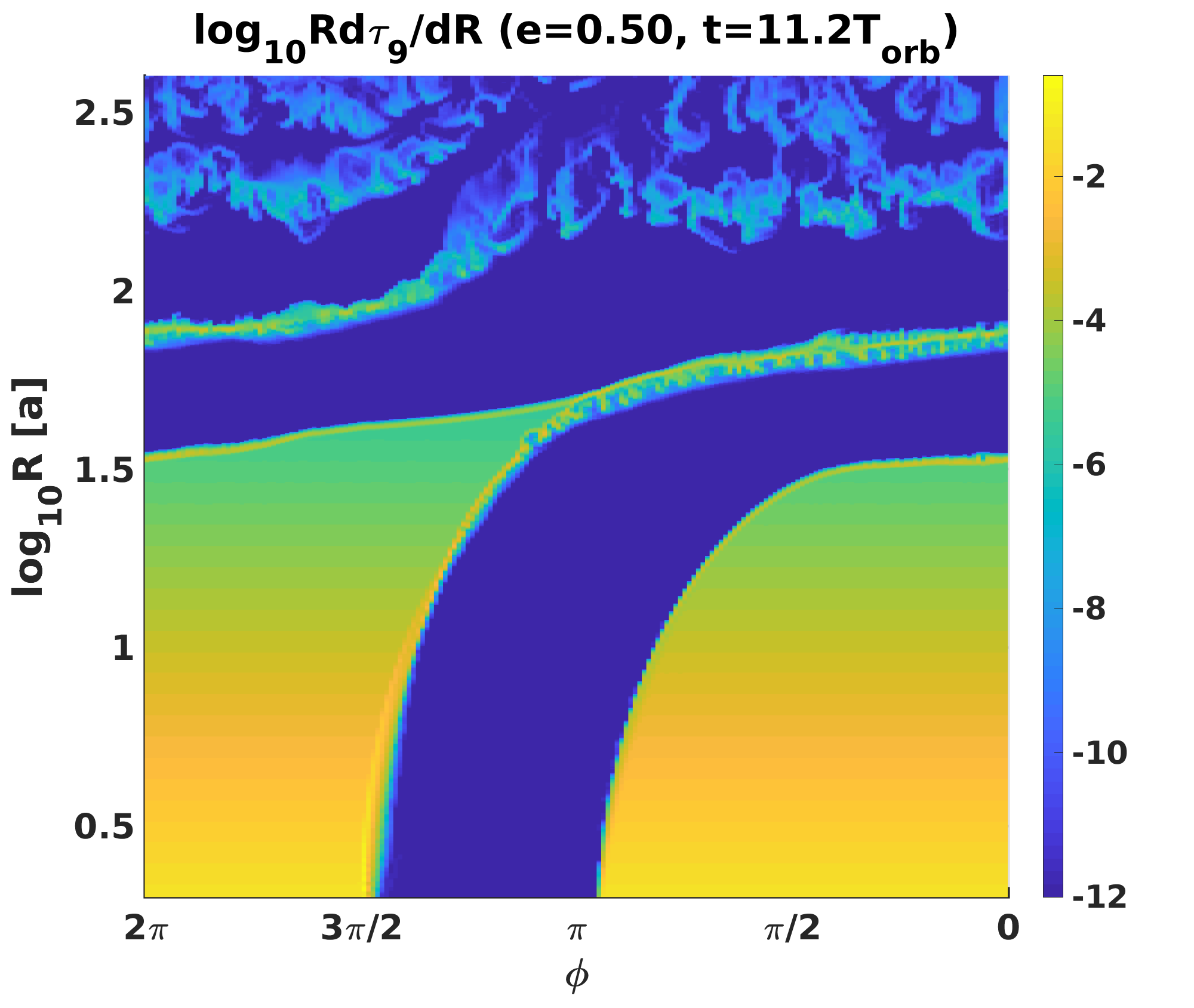}
	\includegraphics[width=\columnwidth]{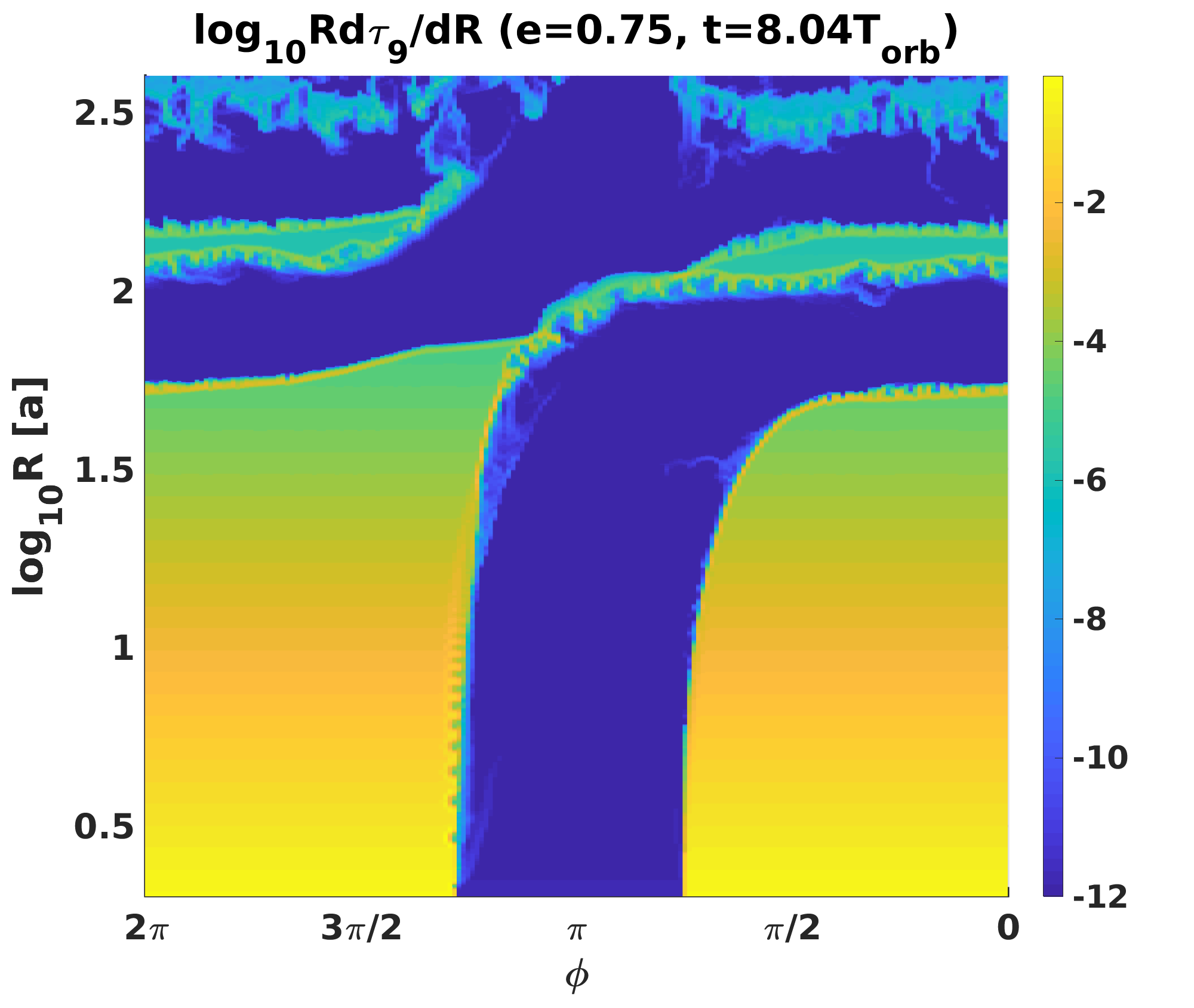}
    \caption{The differential optical depth  of the stellar wind, $R\, \mathrm{d}\tau_9/\mathrm{d}R$, at 1~GHz  is presented for different eccentricities: $e=0.0$ (top left),  $e=0.25$ (top right), $e=0.5$ (bottom left),  $e=0.75$ (bottom right). In all panels we choose orbital phases  soon after the apastron passage. All data corresponds to the orbital plane. The differential optical depth varies with the distance $R$ and the  viewing angle $\phi$ in the orbital plane. }
    \label{fig:dtauphi0}
\end{figure*}

\begin{figure*}
	\includegraphics[width=\columnwidth]{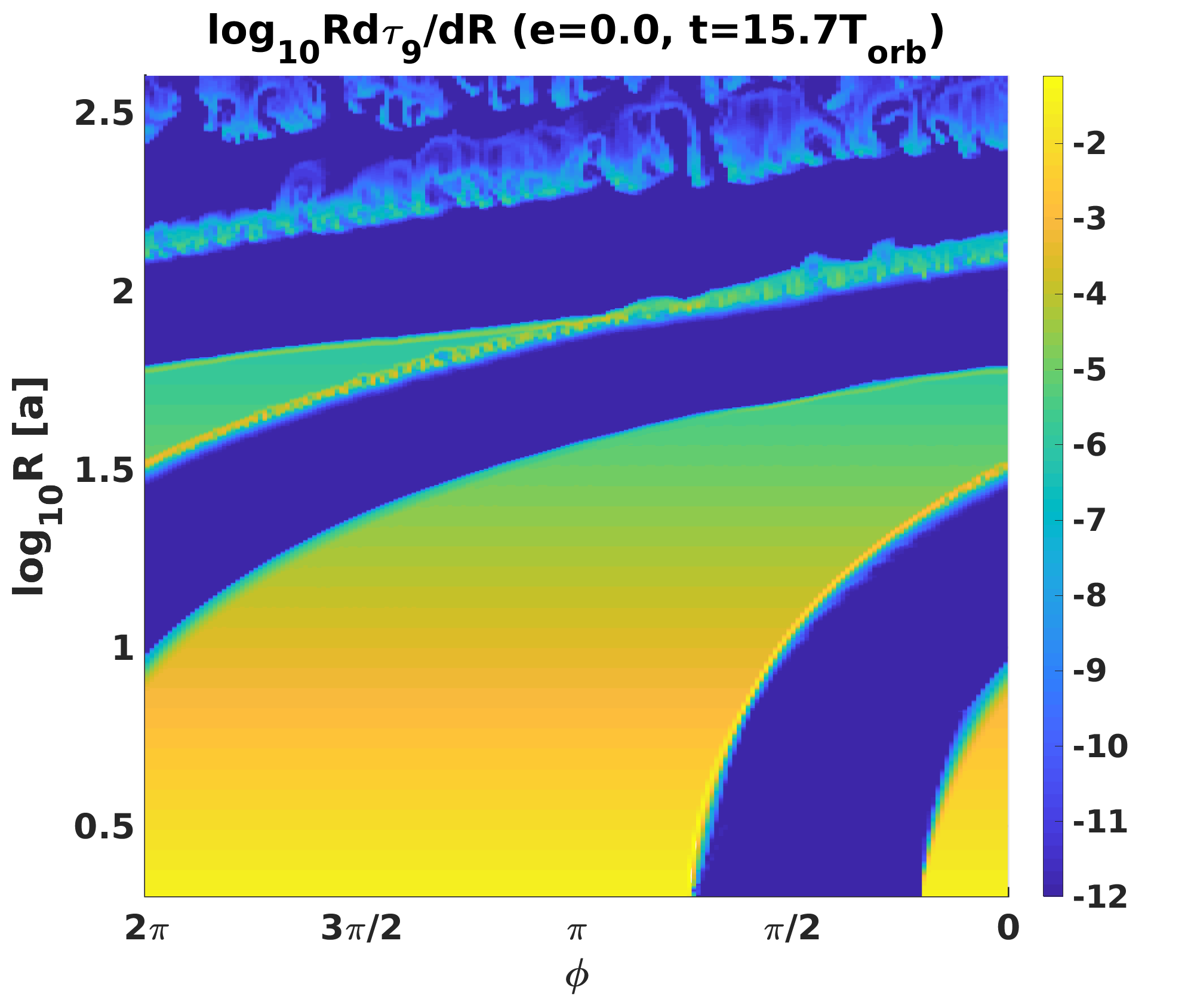}
	\includegraphics[width=\columnwidth]{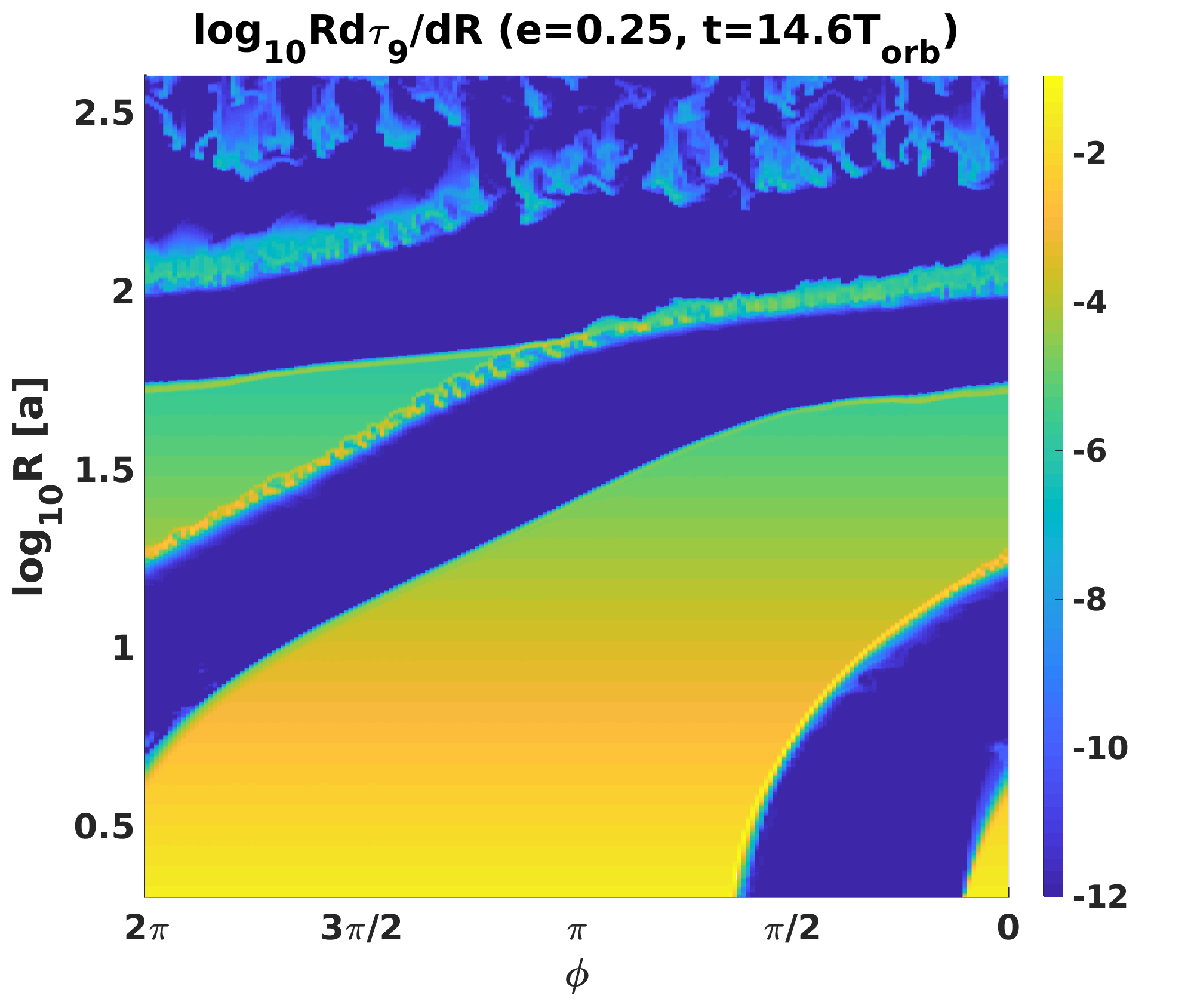}
	\includegraphics[width=\columnwidth]{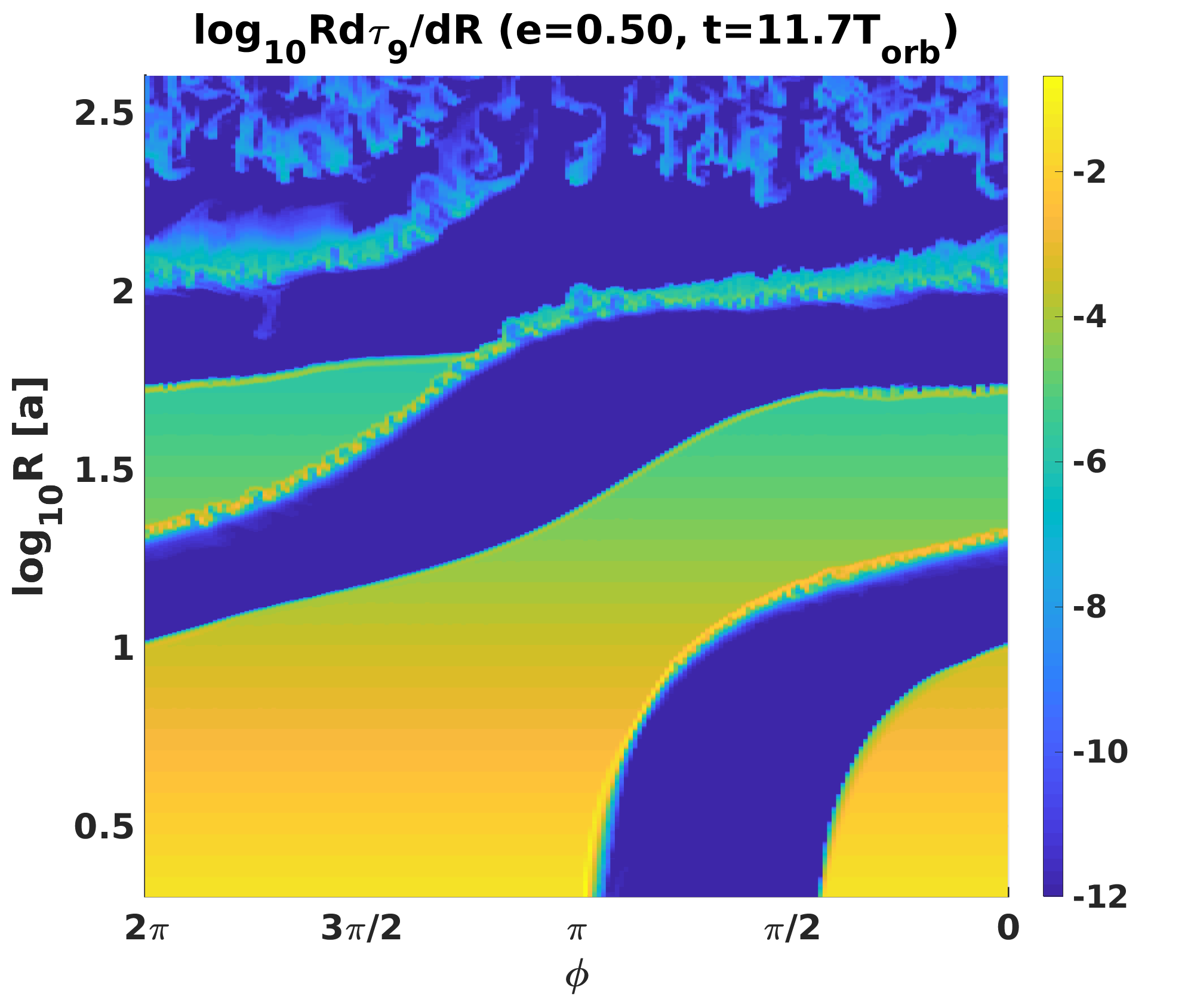}
	\includegraphics[width=\columnwidth]{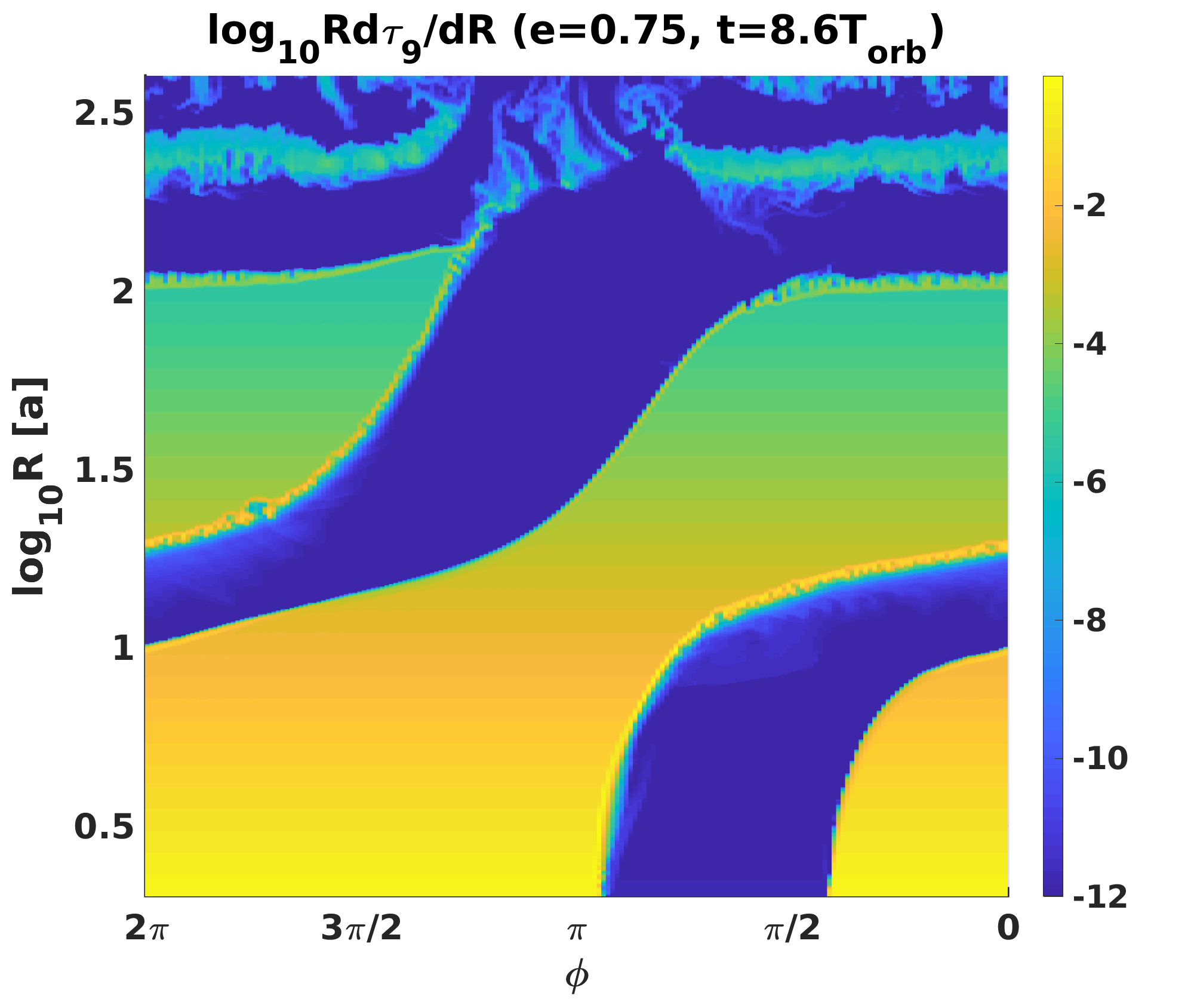}
    \caption{The differential optical depth  of the stellar wind, $R\, \mathrm{d}\tau_9/\mathrm{d}R$, at 1~GHz  is presented for different eccentricities: $e=0.0$ (top left),  $e=0.25$ (top right), $e=0.5$ (bottom left),  $e=0.75$ (bottom right). In all panels we choose orbital phases  soon after the periastron passage. All data corresponds to the orbital plane. The differential optical depth varies with the distance $R$ and the  viewing angle $\phi$ in the orbital plane.
    }
    \label{fig:dtauphi06}
\end{figure*}

\section{The model}
\label{sec:mod}

In this section we present the model used for our calculations.

\subsection{General setup}

In our model we link the burst periodicity of  \frbs{} with  the orbital motion of a magnetar (pulsar) around the center of mass in an eccentric binary system. A schematic view of the system is presented in Fig.~\ref{fig:sketct}. When the compact object is close to the periastron, absorption of a radio signal in the dense stellar wind of the optical companion prevents its detection.  The radio flashes can escape the system only when the magnetar is close to the apastron. The Kepler law dictates how the orbital semi-major axis, $a$, and period, $T_\mathrm{orb}$, are related:
\be
T_\mathrm{orb} = 2\pi \sqrt{ \frac{a^3}{G M_\odot (m_\mathrm{NS}+ m_\mathrm{MS})}}\,.
\ee
{Here $G$ is the Newton constant.}
Thus, we obtain an estimate for the semi-major axis:
\be
a = 4 \times 10^{12} m_{\rm tot,1}^{1/3} T_{\rm orb,6.1}^{2/3} \; {\rm cm},
\label{eq:ac}   
\ee
where $m_\mathrm{NS}$ is \NS's mass and $m_\mathrm{MS}$ is  the primary  mass  in Solar masses and $m_{\rm tot}=(m_\mathrm{NS}+ m_\mathrm{MS})$. Through this paper we adopt the notation $A_x=A/10^x$ and we use Solar mass units, \(M_\odot\), and seconds for normalization of $m_{\rm tot}$  and \(P\), respectively. As it can be seen from eq.~\eqref{eq:ac} the dependence of \(a\) on the uncertain parameter \(m_{\rm tot}\) is quite weak, thus effectively eq.~\eqref{eq:ac} defines the key parameter of the system. The periastron and apastron separations between the stars are \((1-e)a\) and \((1+e)a\), respectively. Here $e$ is the eccentricity of the orbit. We analyse the effect of eccentricity on the absorption. 

\subsection{Winds and shocks}

The physics of interaction of the pulsar and stellar winds in a binary system harboring a powerful pulsar and a massive optical companion was widely studied in the literature by means of  2D and 3D numerical simulations \citep[for details see ][]{bog08,bog12,bbkp12,2015A&A...577A..89B,dub15,2019MNRAS.490.3601B}. Under the assumption of isotropic winds, geometry in the apex region is determined by a single parameter --  the ratio of the winds momentum flux:
\be
\eta= \frac{L_{\rm pw}}{\dot{M}_\mathrm{w} c \varv_\mathrm{w}}\,,
\ee
{where $c$ is the speed of light, \(L_{\rm pw}\), \(\dot{M}_\mathrm{w}\), and \(\varv_\mathrm{w}\) are power of the pulsar wind, optical companion mass-loss rate, and
speed of the stellar wind in the collision region, respectively.} For realization of our scenario, we need to assume that the parameter \(\eta\)  is quite small and we adopt \(\eta\simeq0.1\), see also below Sec. 2.4. If the pulsar wind is powerful, \(\eta\gtrsim1\), then it removes the stellar wind from the binary system, making the free-free absorption less efficient \citep{2020ApJ...893L..39L}. 

Modeling of the winds interaction region on a scale exceeding
the separation between components of the binary, typically requires high-resolution 3D simulations to account accurately for the
orbital parameters and relative motion of the stars.  \citet{2016MNRAS.456L..64B} proposed a computationally efficient method in which one truncates the central part and uses low vertical resolution (relative to orbital plane of the system). This simplified
approach allows to model wind evolution at a distance of several hundreds of the orbital
separation  relatively easy. In particular, it is possible to study formation of a wind mixing zone for highly eccentric binaries.

The binary motion results in formation of a spiral structure of shocked pulsar wind surrounded by a more dense and slow stellar wind \citep{2011A&A...535A..20B,2015A&A...577A..89B}. In the case of large eccentricity of the orbit ($e\gtrsim 0.75$) the interaction of the stellar and pulsar winds forms an asymmetrical spiral structure \citep[][]{2021Univ....7..277B}. 

In the case of an eccentric orbit, the `back termination shock' which is caused by the Coriolis force \citep{bbkp12} 
is situated at a distance of several tens of the orbital separation from the neutron star (see in Fig.~\ref{fig:Tdist} for illustration):
\be
R_\mathrm{s} 
=\chi_a a 
\approx 1.2\times 10^{14} m_{\rm tot,1}^{1/3} \chi_{a, 1.5} \rm\,cm\,.
\label{eq:Rs}
\ee
here we parameterize  location of the shock on the cavity side (i.e., opposite to the optical star)  -- the Coriolis shock, -- as  $\chi_a a$. 
 Numerical simulations \citep[][]{2021Univ....7..277B} suggest that for eccentric orbits \(\chi_a\) should be quite large. Thus, in numerical estimates we normalize it as $\chi_a = 10^{1.5} \chi_{a,1.5}$.

We should note that in the high mass binary system \psr{}  pulsations are observable if the pulsar is at a distance larger than $1.6\mbox{ AU}\approx 2.5\times10^{13}$~cm \citep[][]{2011ApJ...732L..10M}. So,  we can expect transparency of the stellar wind for a significant part of the `pulsar wind tail'.  

\subsection{Transparency for radio emission}

The free-free optical depth accumulated from the location of the Coriolis shock to infinity should be at most of order unity  to allow radio waves to reach an observer.
We use the free-free absorption coefficient  $\kappa_\mathrm{ff}$ \citep{1999acfp.book.....L} and apply the procedure for the optical depth calculation from \citet{2020ApJ...893L..39L}. We find:
\be
\tau _\mathrm{ff}  \sim   4\times10^{-3} \; \chi_{a,1.5}^{-3}   \dot{M}_{-7.5}^2  T_4^{-1.35} \nu_\mathrm{1 GHz}^{-2.1} m_{\rm tot,1} \varv_{w,8.5}^{-2}\,.
\label{Mdot}
\ee
Here  $T_4=T/(10^4{\rm\,K})$ is the  temperature of the absorbing plasma. 
For the adopted parameters, a radio source located at a distance of \(\sim a\) should be significantly attenuated. However, for \(\chi_a\sim30\) the stellar wind becomes transparent.    

The stellar wind should be powerful enough to dominate the pulsar wind. On the other hand, if the stellar wind is too powerful then the free-free absorption is large enough to attenuate the signal emitted also at a large distance from the optical companion. Winds expected from early 
B-type stars  and late O-type stars seem to fit the required mass-loss range.  Late B-type stars have insufficiently strong winds, while winds of earlier O-type stars remain optically thick up to large distances \citep{2001A&A...369..574V}.

\begin{figure*}
	\includegraphics[width=\columnwidth]{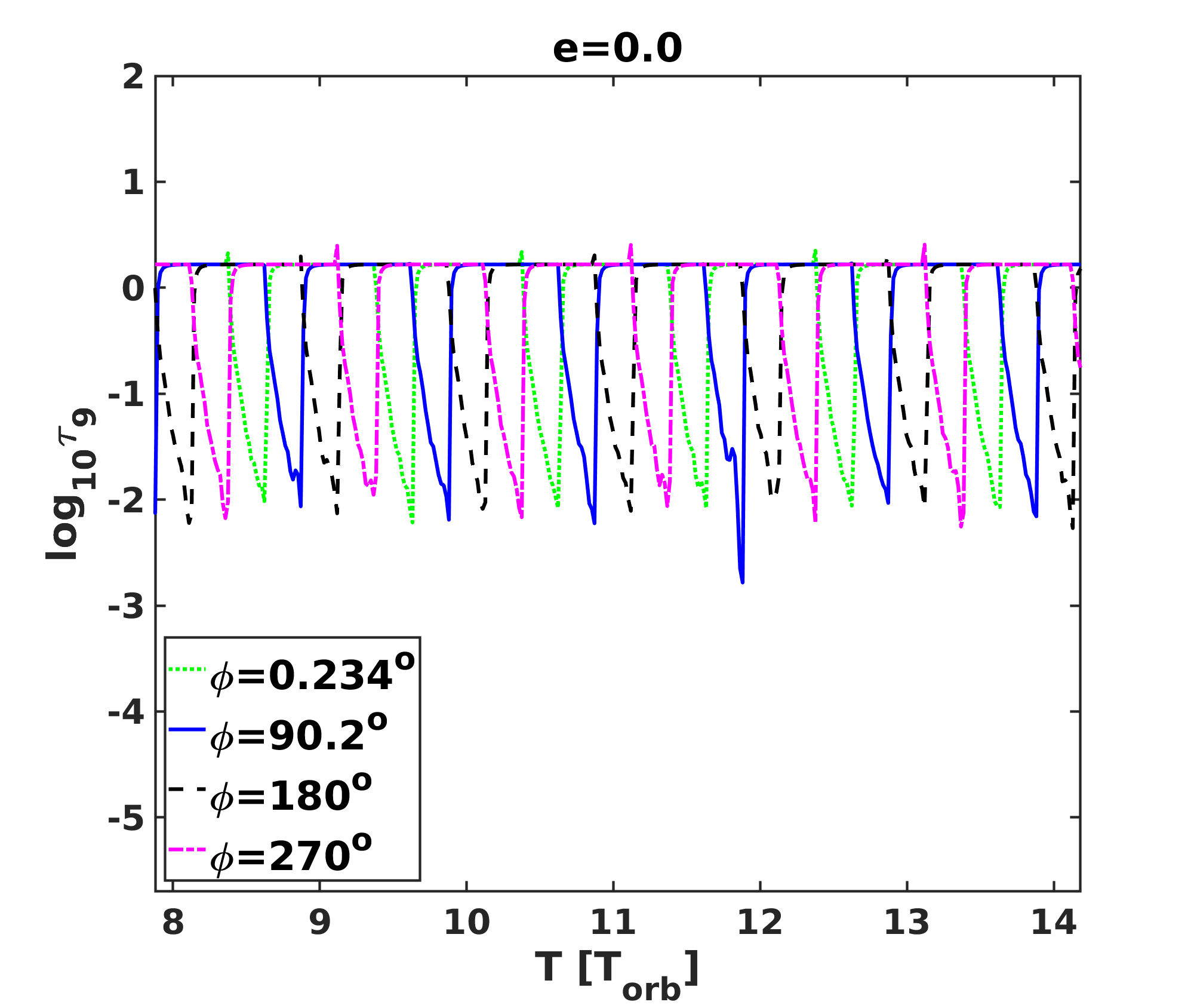}
	\includegraphics[width=\columnwidth]{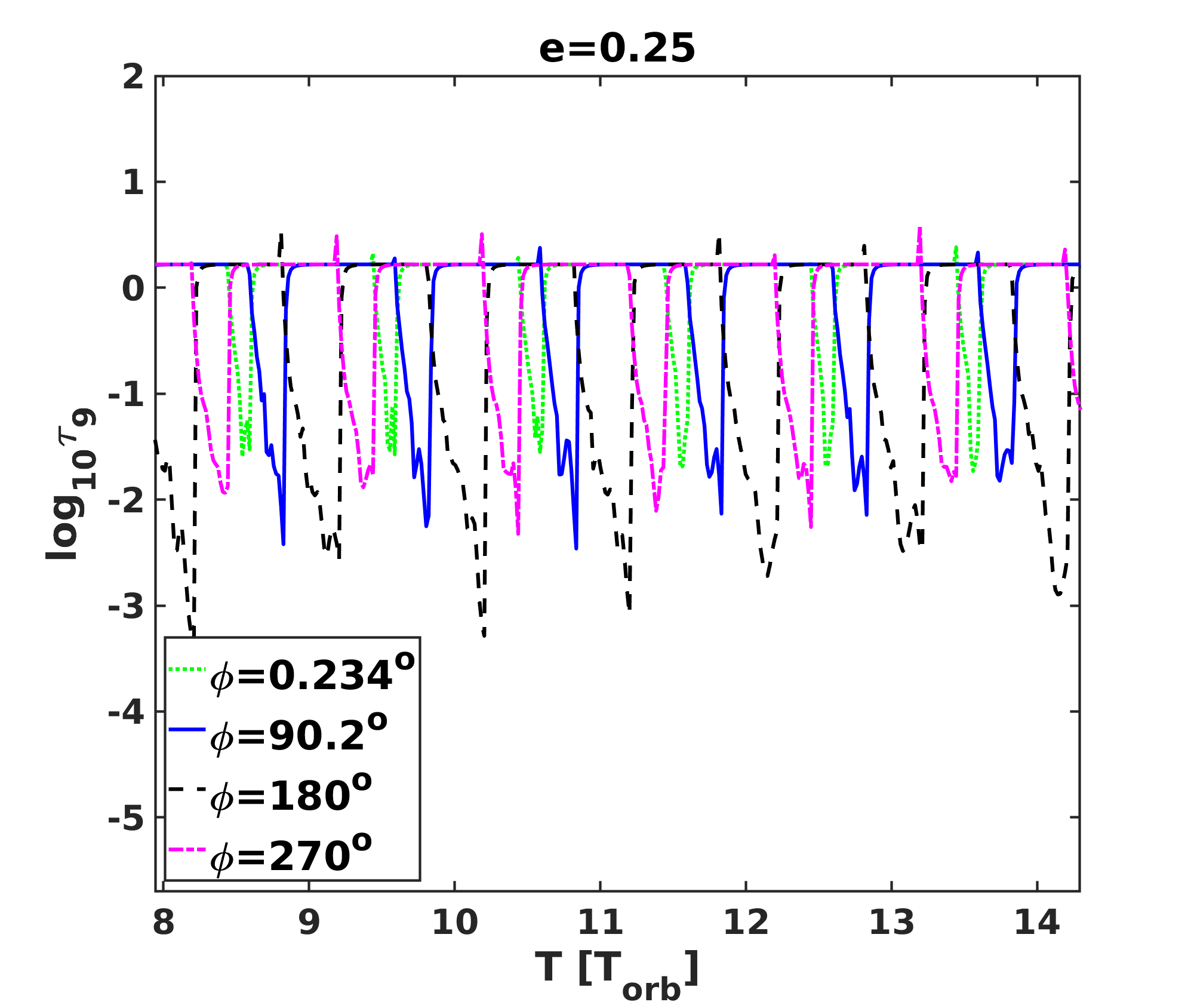}
	\includegraphics[width=\columnwidth]{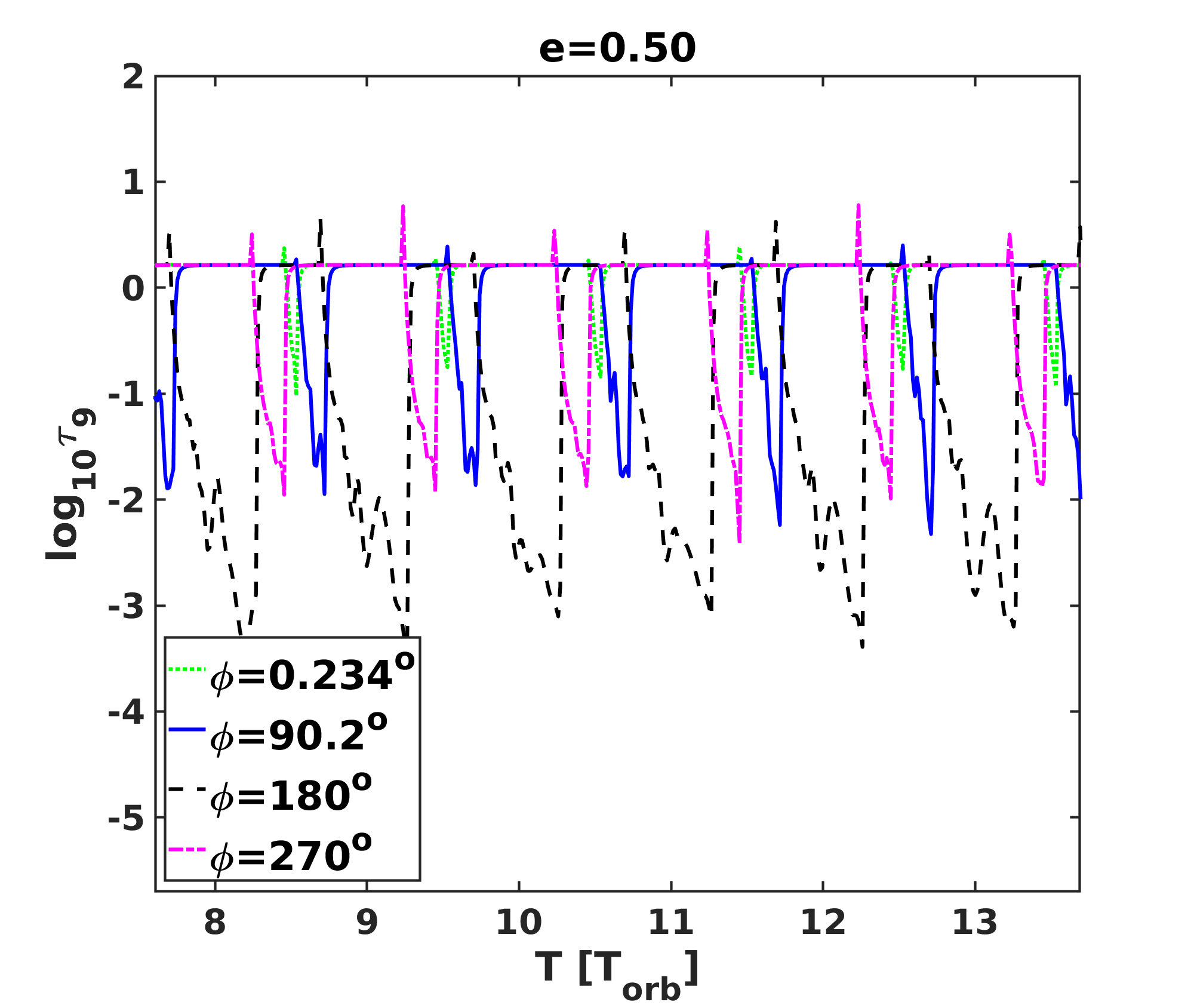}
	\includegraphics[width=\columnwidth]{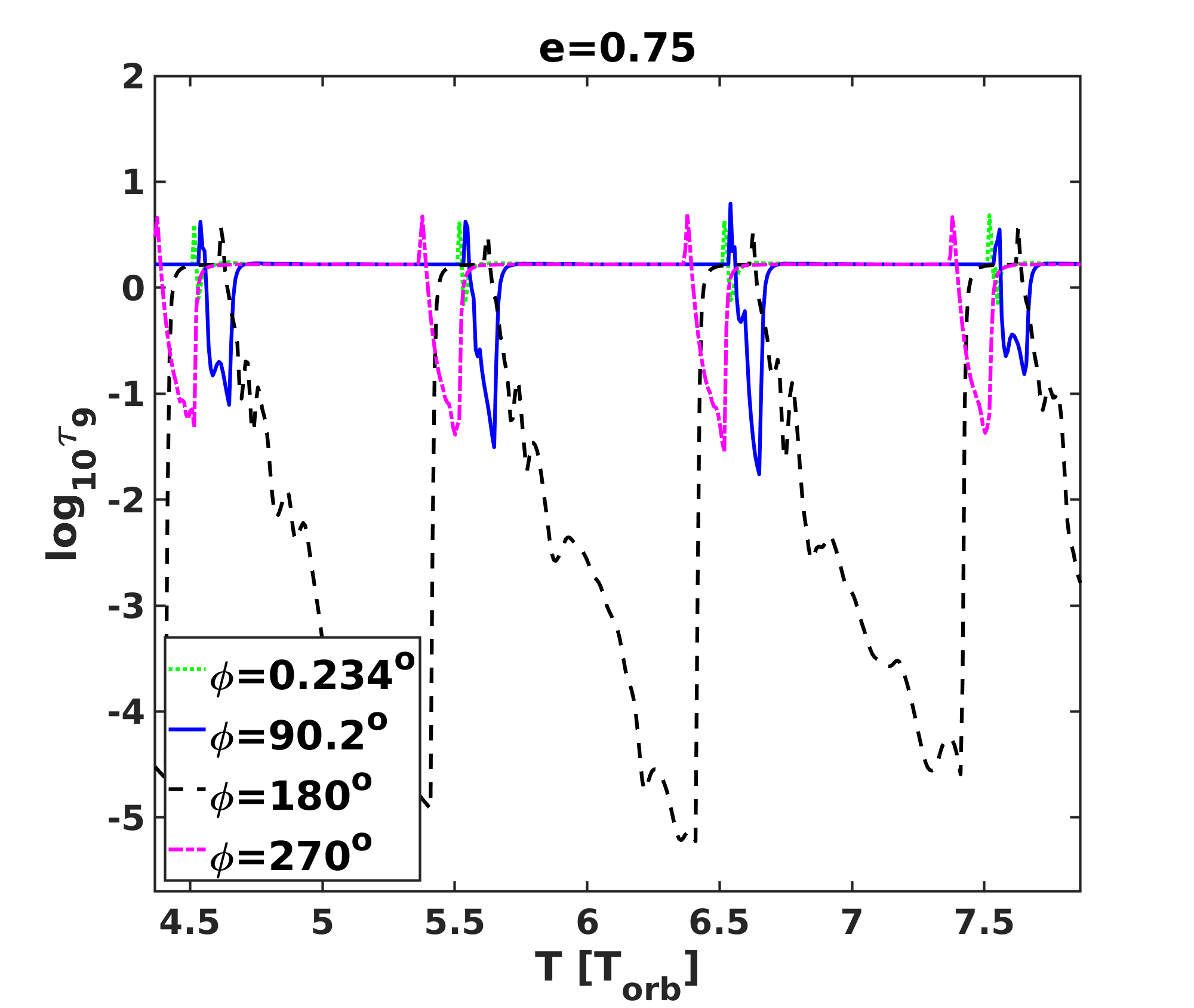}
    \caption{Changes of the  optical depth of the stellar wind $\tau_9$ (at 1~GHz) during several orbital periods as viewed by observers at different viewing angles $\phi$. Different panels are plotted for different eccentricities: $e=0.0$ (top-left panel), $e=0.25$ (top-right panel), $e=0.50$ (bottom-left panel), $e=0.75$ (bottom-right panel).  }
    \label{fig:taut}
\end{figure*}

\begin{figure*}
	\includegraphics[width=\columnwidth]{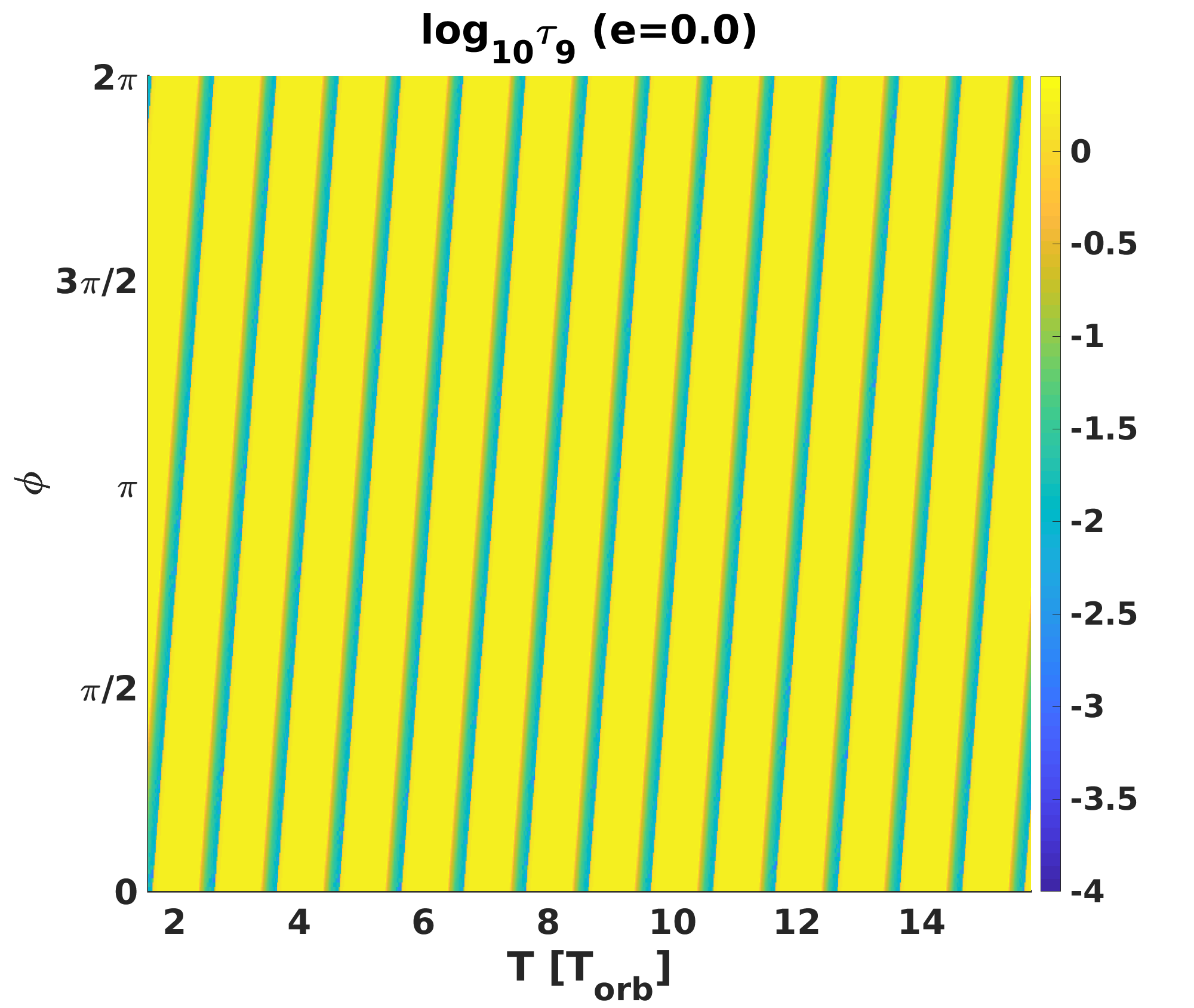}
	\includegraphics[width=\columnwidth]{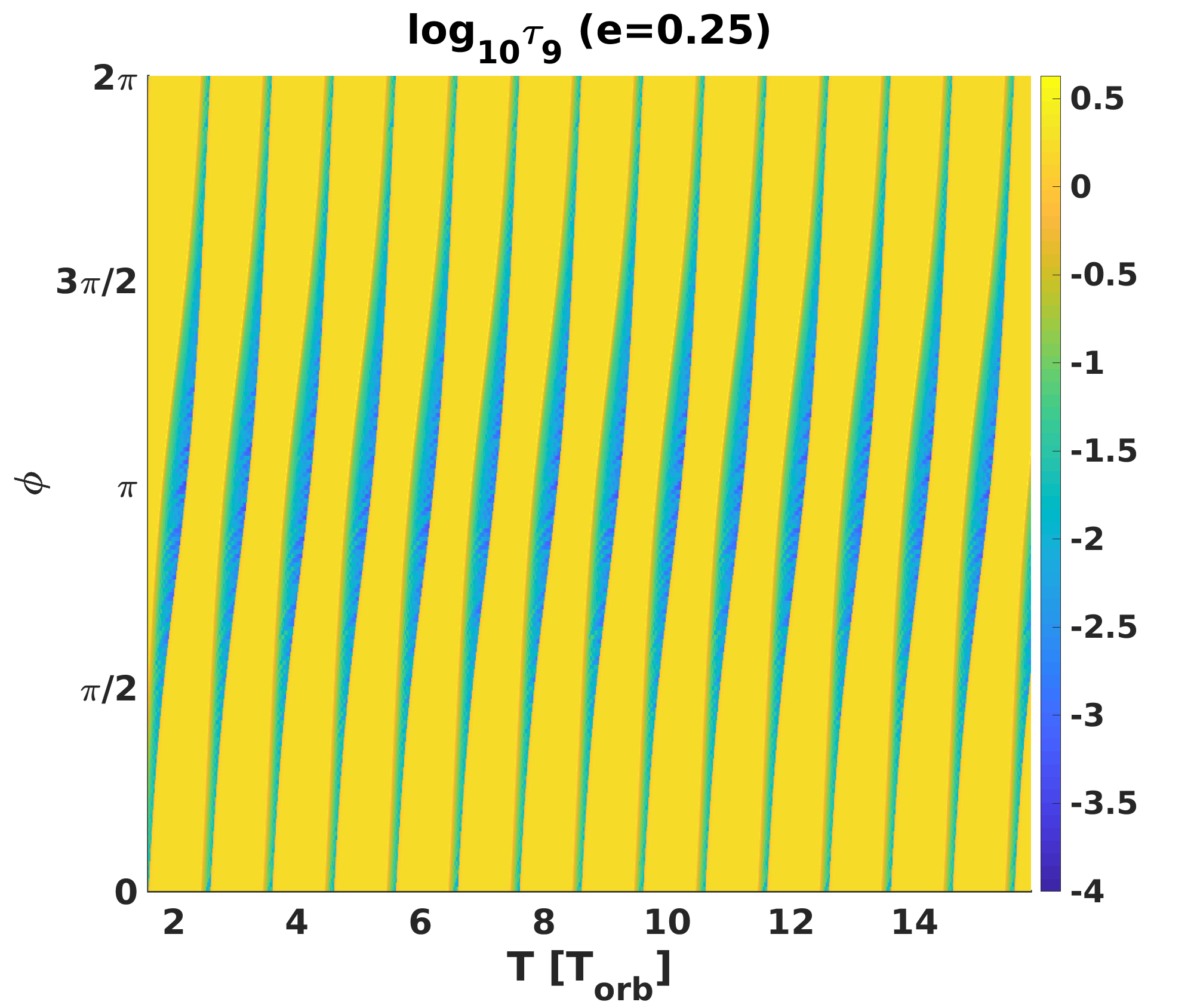}
	\includegraphics[width=\columnwidth]{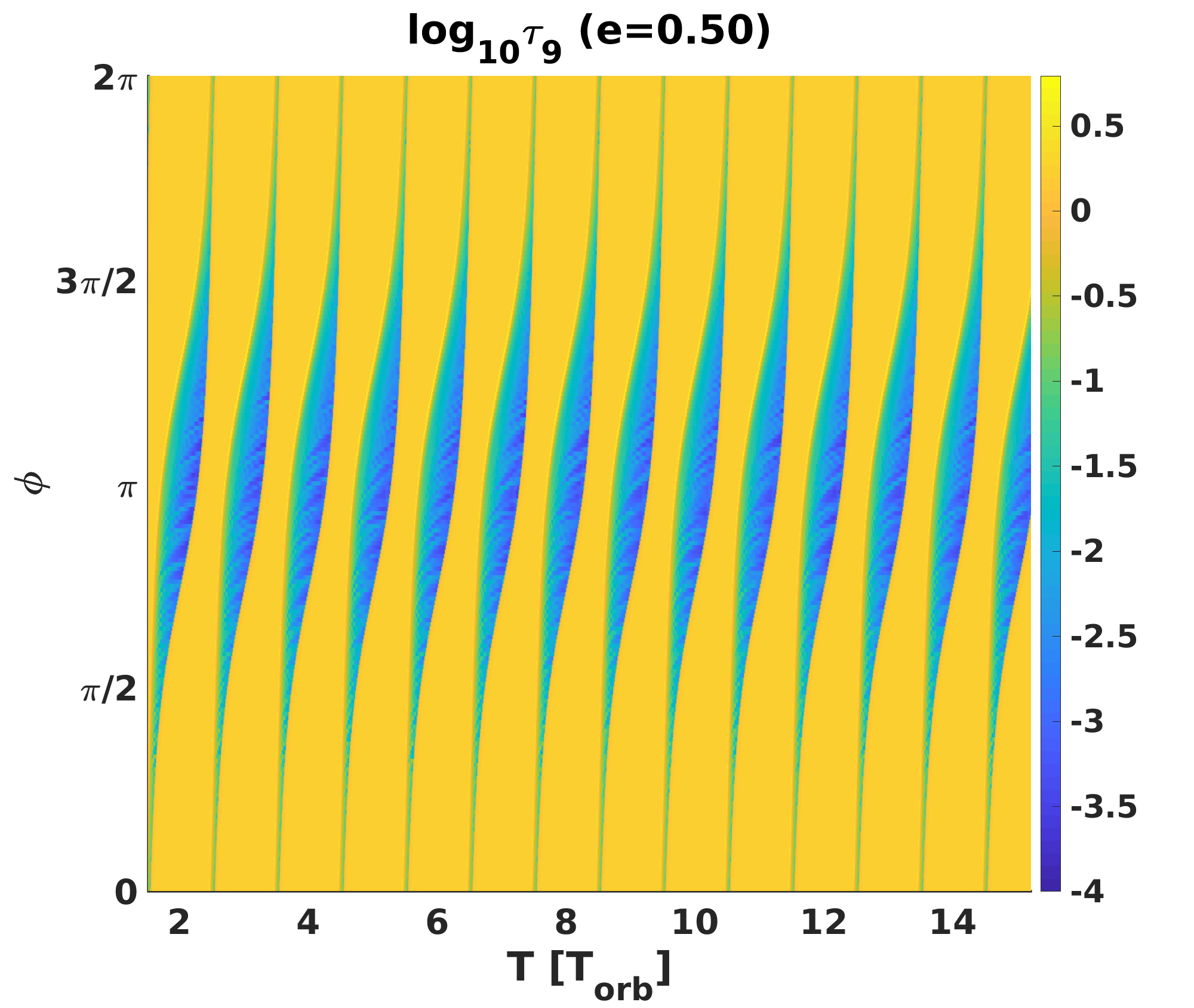}
	\includegraphics[width=\columnwidth]{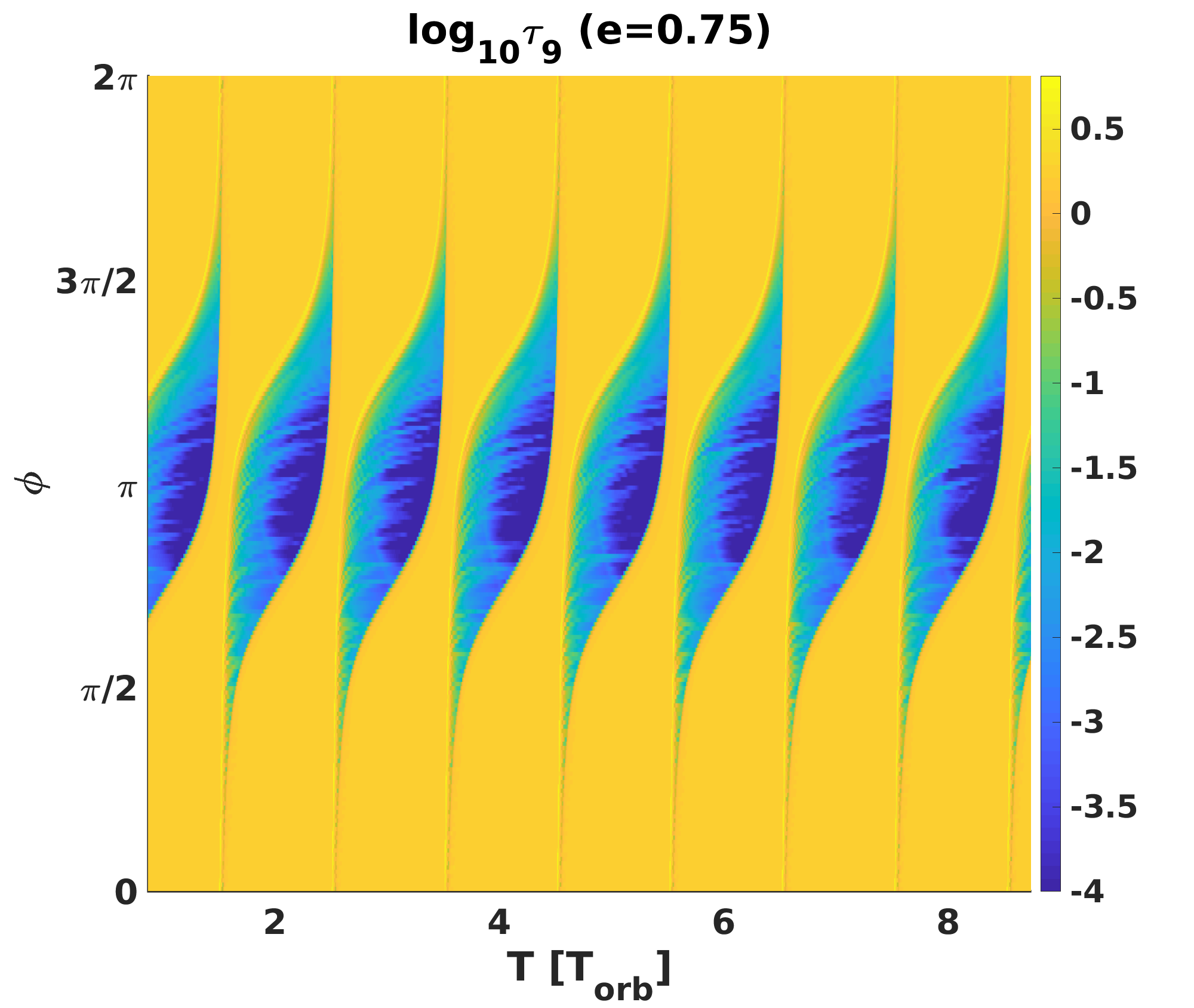}
    \caption{The optical depth of the stellar wind $\tau_9$ (at 1~GHz) for different eccentricities: $e=0.0$ (top-left panel), $e=0.25$ (top-right panel), $e=0.50$ (bottom-left panel), $e=0.75$ (bottom-right panel). The plot allows to have an idea, how the optical depth changes during several orbital periods for the whole range of the viewing angle.
    }
    \label{fig:tauphitdist}
\end{figure*}

\begin{figure}
	\includegraphics[width=\columnwidth]{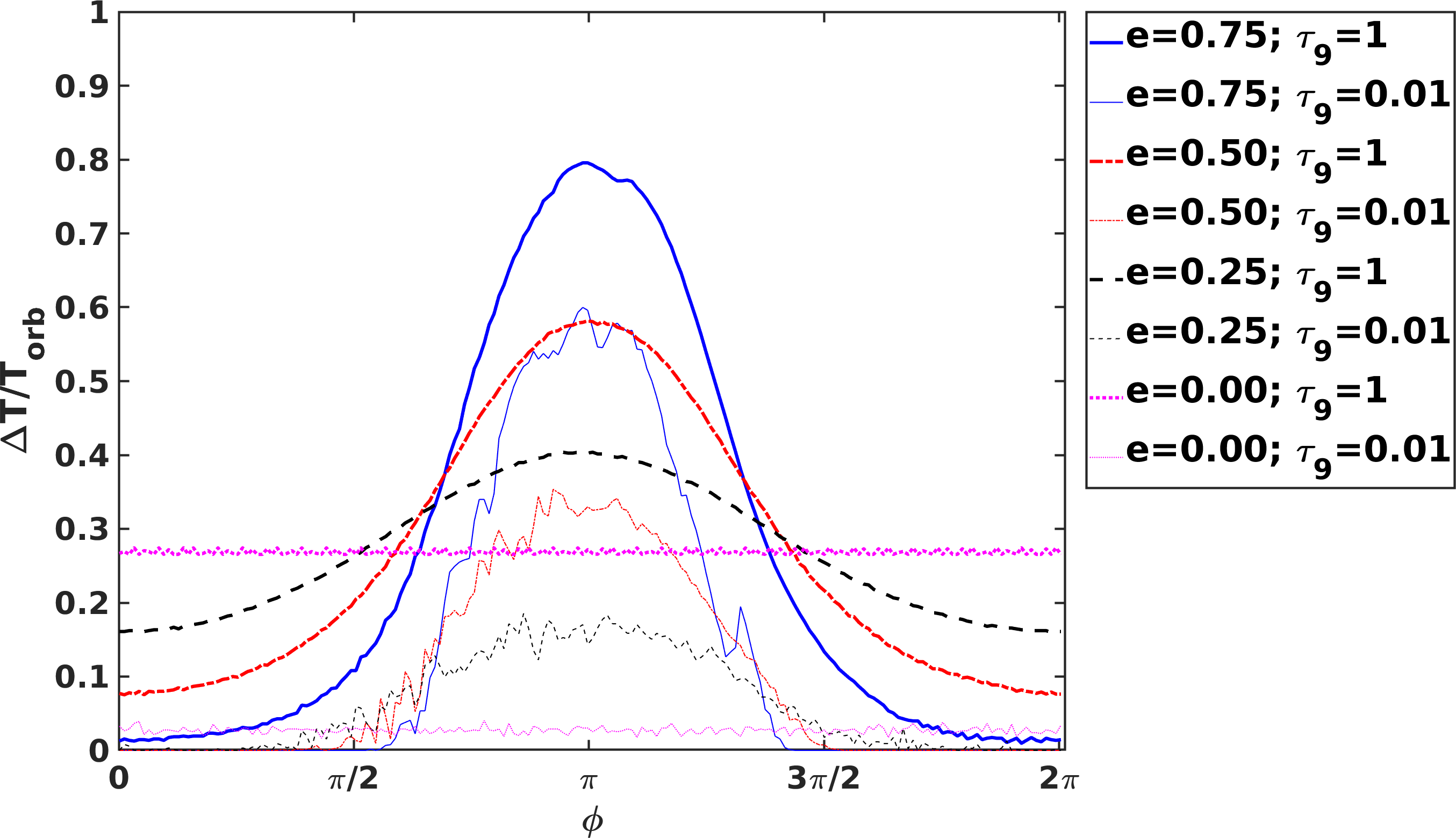}
    \caption{Curves represent a fraction of the orbital period when $\tau_9$ at 1~GHz is smaller than the given value (1 -- thick lines; 0.01 -- thin lines) versus the viewing angle $\phi$. Different curves correspond to: $e=0.0$ (magenta dotted lines); $e=0.25$ (black dashed lines); $e=0.5$ (red dot-dashed lines), and  $e=0.75$ (blue solid lines). Here we set $\dot{M} = 3\times10^{-8} M_\odot$~yr$^{-1}$.}
    \label{fig:taudTT}
\end{figure}

\subsection{Numerical model}
\label{sec:nmodel}

For our numerical simulations we use the {\textsc PLUTO} code \citep{2007ApJS..170..228M} in 3D spherical coordinates $(R,\theta,\phi)$ with resolution $(768,3,768)$. So, the resolution in the $\theta$ direction is very low. We call such approach (3-1)D \citep[][]{2016MNRAS.456L..64B}. This approach shows good agreement on  large scales with 3D relativistic hydrodynamic (RHD) results. Finally, the (3-1)D approach allows us to perform long simulations in a wide range of radial distances. 

General hydrodynamical properties of the model 
were presented in the paper \citep[][]{2021Univ....7..277B}. 
We use spherical coordinates centered on the center mass of the system with $R$ being the distance from the coordinate center and two angles -- $\theta$ and $\phi$. The later angle is in the orbital plane, where $\phi=0$ corresponds to the direction towards the periastron.
Coordinates can vary in the following ranges: $R=(1-e^2)\times[2\,a,1000\,a]$, $\theta = [\pi/4,3\pi/4]$, and $\phi = [0,2\pi]$.  The orbital period of the system is assumed to be equal to $T_\mathrm{orb}=16$~days which corresponds to  $a=6\times10^{12}$~cm.  We study four cases with different  orbital eccentricities: $e=0.0$, $e=0.25$, $e=0.5$, and $e=0.75$. The normal star mass loss rate is assumed to be equal to $\dot{M}=3\times 10^{-8} \; M_\odot $~yr$^{-1}$. The winds thrust ratio is $\eta = L_{\rm pw}/\dot{M}_\mathrm{w} \varv_\mathrm{w} c = 0.1$  in agreement with \cite{2020ApJ...893L..39L}.  This corresponds to the half-opening angle of the pulsar wind tail $\theta_p = 50^\circ$. 
The value $\eta=0.1$ is relatively high for a magnetar wind, but it can be boosted by consequence of magnetar flares \citep[see discussion in ][]{2022ApJ...927....2K}. Note, that the winds thrust ratio is not changing along the orbit.

\section{Results}
\label{sec:results}

At first, for an illustration we present the general structure of the flow obtained in the (3-1)D relativistic hydrodynamical simulations, see Fig.~\ref{fig:Tdist}. In this plot we show the temperature distribution in a hydrodynamical flow for the high eccentricity case $e=0.75$. The left panel shows the flow structure near the periastron phase, and the right panel -- near the apastron. We see significant evolution of the flow depending on the orbital phase. Here we do not show hydrodynamical properties of the flow in details \citep[details can be found in the paper ][]{2021Univ....7..277B}.

The distance where the Coriolis turnover tail is formed can be estimated by a simple formula \citep[][]{2011A&A...535A..20B,2015A&A...577A..89B}:
\be
    \chi_0 = \frac{3 v_\mathrm{w}}{2\Omega}\eta^{1/2},
\label{eq:chicor}
\ee
{here $\Omega$ is the orbital angular velocity.}
If we take into account the orbital eccentricity then the maximum turnover radius becomes
\be
    \chi_{\rm max} = \chi_0\sqrt{\frac{(1+e)^3}{1-e}}.
\label{eq:chimax}
\ee
This value for $e=[0.0; \; 0.25;\; 0.5; \; 0.75 ]$ is equal to $\chi_{\rm max}/a = [8;\; 13;\; 21;\; 38]$. The last one is in a good agreement with eq.~(\ref{eq:Rs}) and Fig.~\ref{fig:RTphidist}.

\begin{figure*}
	\includegraphics[width=\columnwidth]{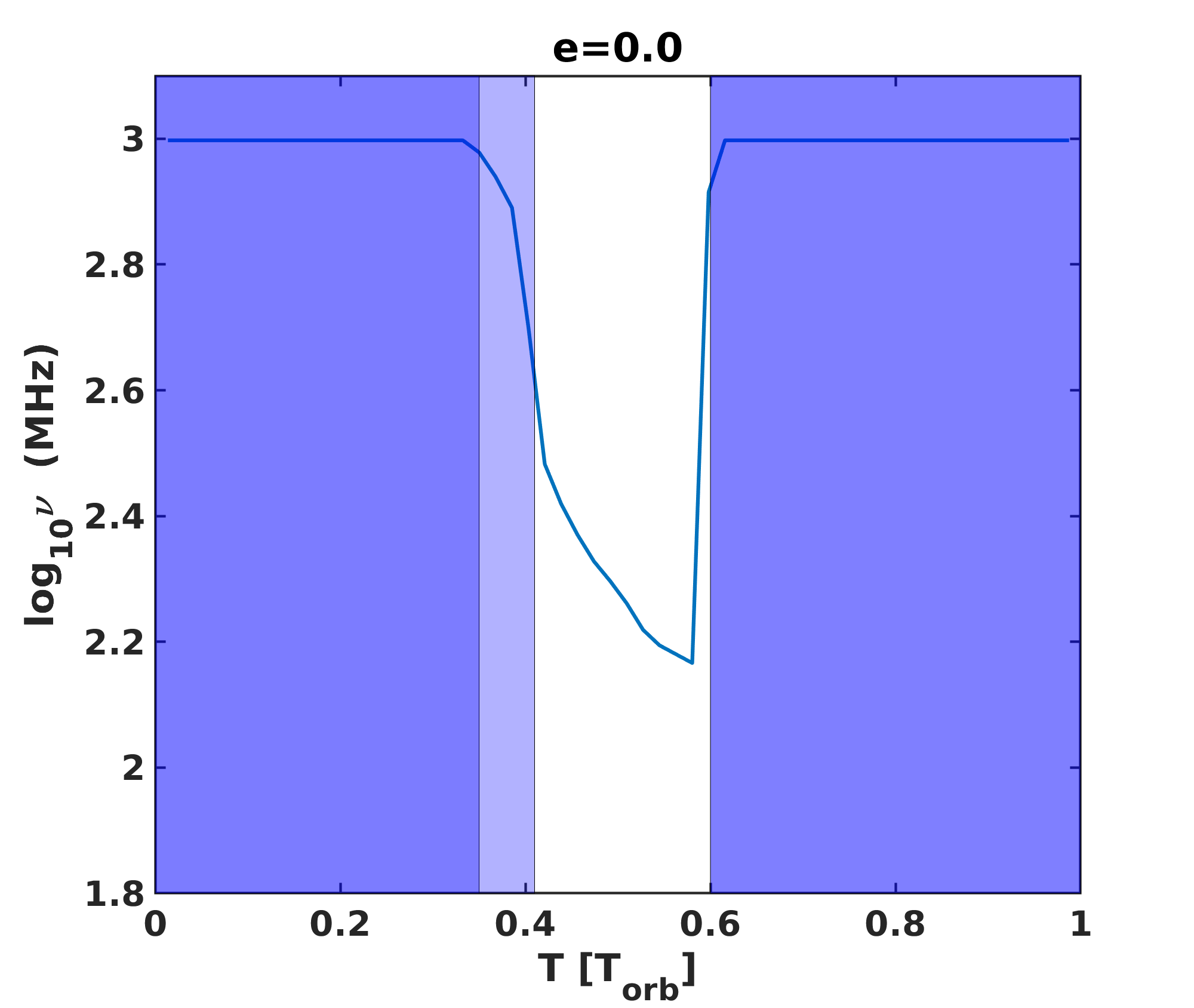}
	\includegraphics[width=\columnwidth]{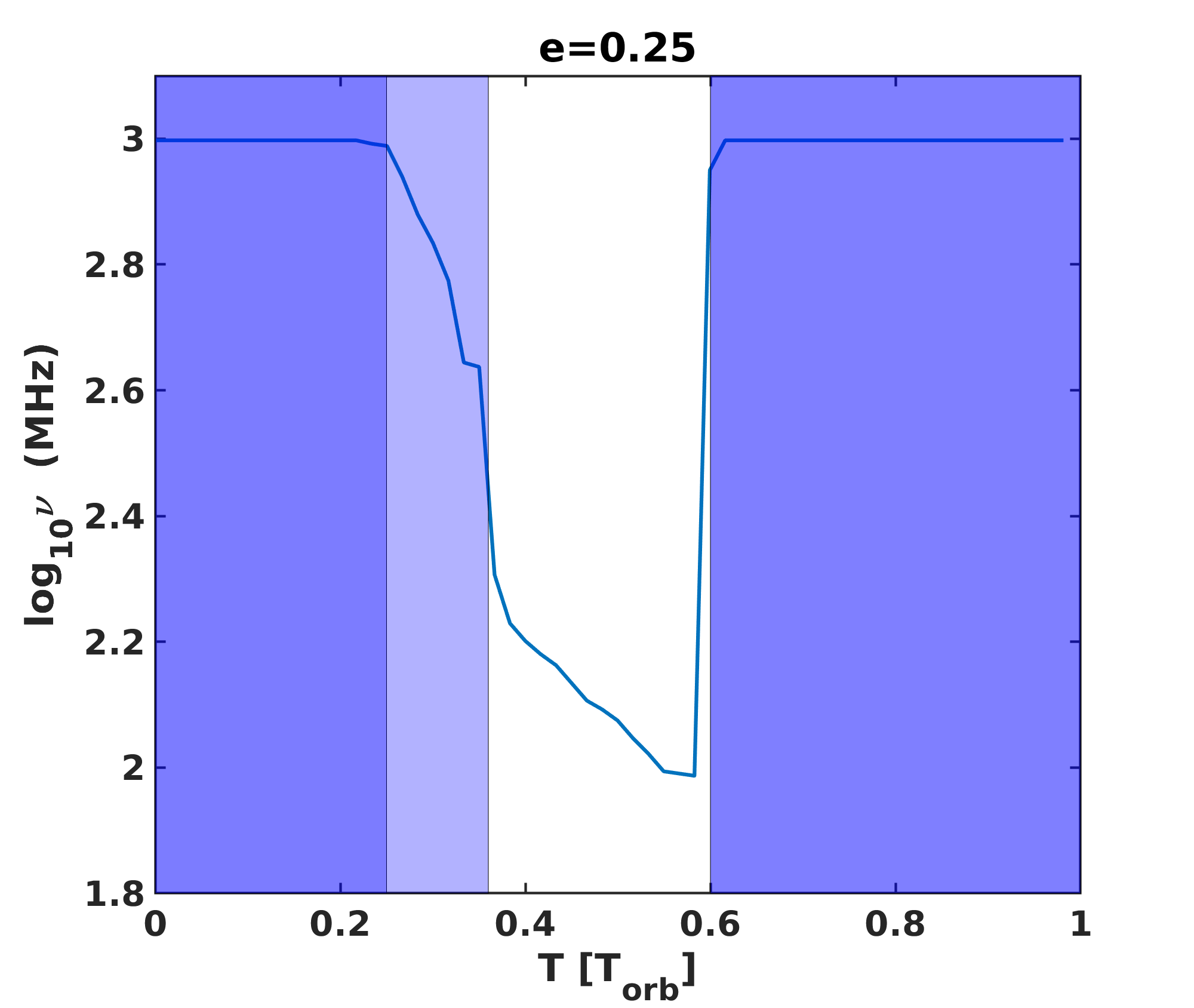}
	\includegraphics[width=\columnwidth]{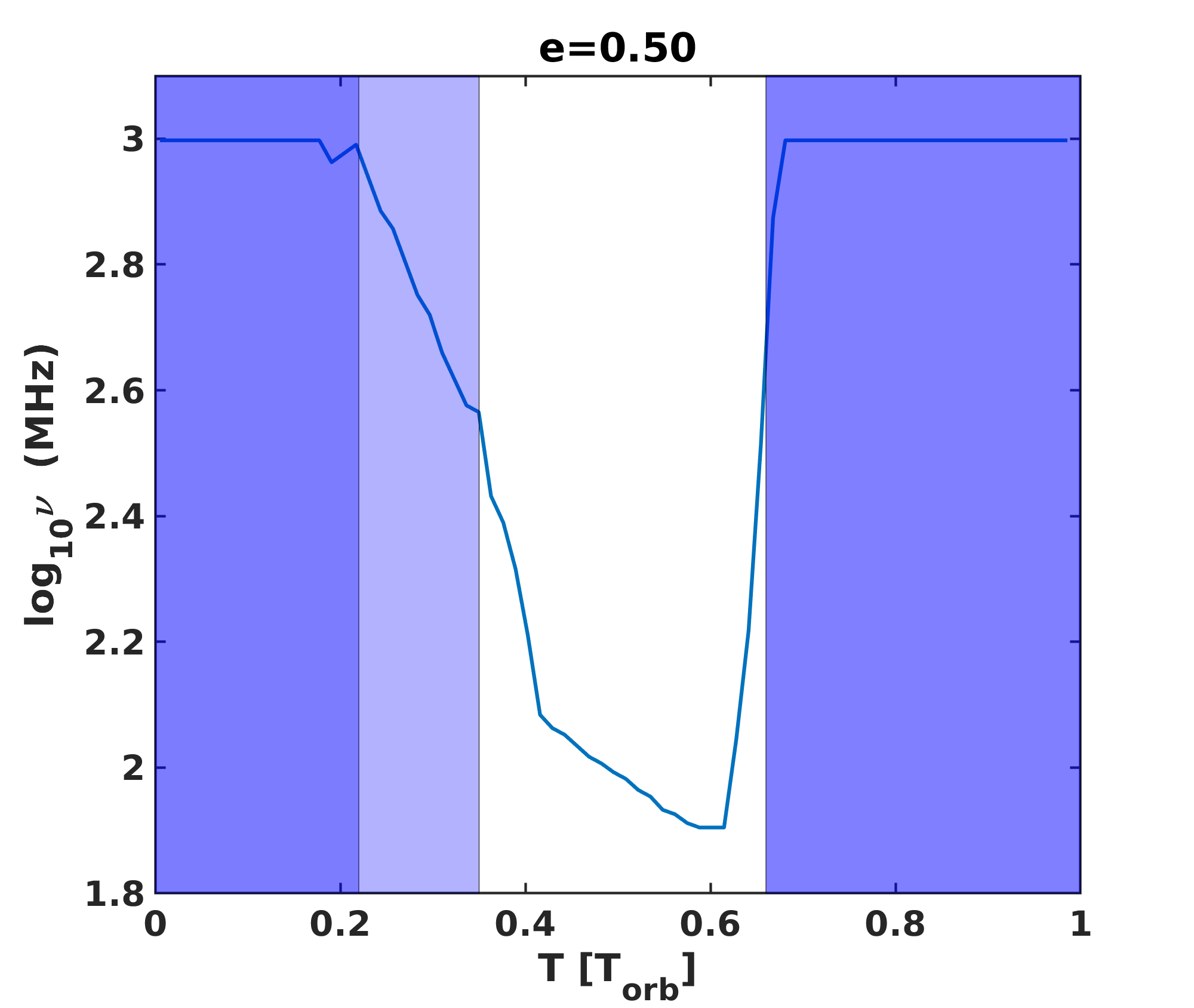}
	\includegraphics[width=\columnwidth]{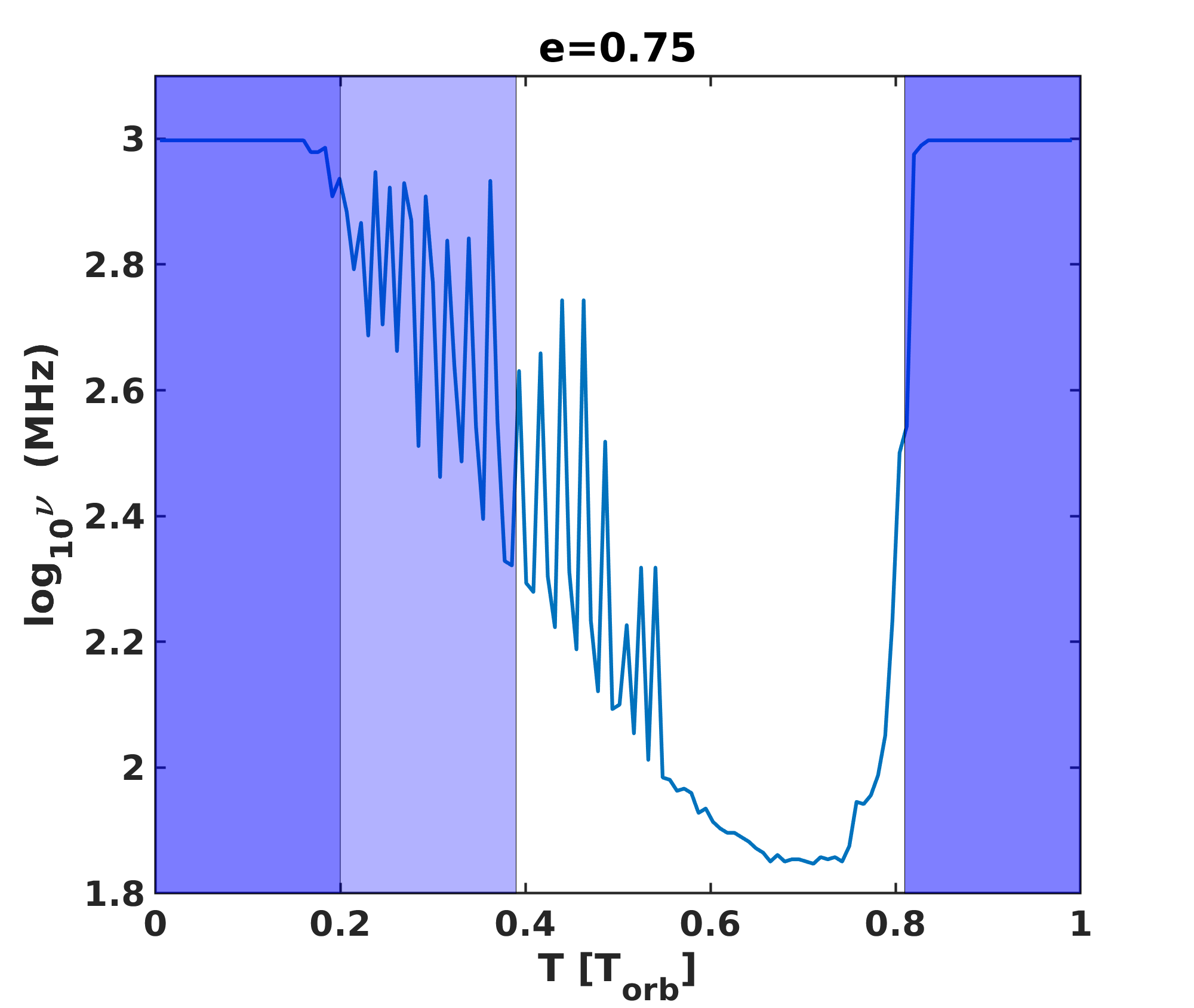}
    \caption{An example of behaviour of the  typical frequency of the radio emission given by eq.~(\ref{eq:omaxrs}) during an orbital period as observed from the viewing angle $\phi=3.57$ radians. 
    All curves are calculated for the total luminosity of a magnetar flare $L_{\rm fl}=4\times10^{42}$~erg~s$^{-1}$ and $\Gamma_{\rm fl} =10^3$. 
    Different panels correspond to the following eccentricities: $e=0.0$ (top-left panel), $e=0.25$ (top-right panel), $e=0.50$ (bottom-left panel), $e=0.75$ (bottom-right panel). {Note, that oscillations visible in the bottom right panel in the descending branch of the curve ($e=0.75$) might be of a numerical origin. Still, the lower envelope curve represents the  trend in the evolution of the typical frequency.} For clarity, we indicate  by a darker colour the zones where the  optical depth at 1 GHz $\tau_9$ is larger than unity and by a lighter colour -- the zones where the optical depth at 100 MHz $\tau_8$ is above unity. 
     White areas in each panel correspond to transparent zones. 
    }
    \label{fig:nutdist}
\end{figure*}

\subsection{Origin of periodicity}

The key point of the hypothesis which interprets  the periodicity of the \frbs{} as an orbital period of a binary, is absorption of the radio emission in the stellar wind \citep[][]{2020ApJ...889..135L}. To illustrate properties of the absorption towards an observer in the matter of the flow, we plot the differential optical depth  at 1 GHz, $R\frac{\mathrm{d}\tau_9}{\mathrm{d}R}$, where $\mathrm{d}\tau_9 = n \kappa_{ff,9} \mathrm{d}R$, in the orbital plane  versus the viewing angle $\phi$ for various eccentricities.  The plot for a NS near the apastron phase is given in Fig.~\ref{fig:dtauphi0}.  In Fig.~\ref{fig:dtauphi06} we show a similar  plot for a NS just after the periastron passage. The length of the pulsar tail at the apastron phase grows with the orbital eccentricity as follows from eq.~(\ref{eq:chimax}) and decreases around the periastron.   

Fig.~\ref{fig:taut} illustrates how the optical depth towards a remote observer in the orbital plane depends on the orbital phase for different eccentricities. Dependence of the optical depth on the viewing angle and orbital phase is shown in Fig.~\ref{fig:tauphitdist}.

For different eccentricities and viewing angles we can obtain various duration of the period when the radio emission can reach  the observer. 
In  Fig.~\ref{fig:taudTT}  we plot dependence of the width of the optically thin window on the viewing angle for several values of the eccentricity.  Naturally, larger $e$ -- larger is the fraction of the orbital period when an observer in the direction close to apastron can detect the radio emission. For observers on the opposite side the situation is reversed. 


\subsection{Radio emission from relativistic shocks }
\label{sec:Emission}

%
Several variants of  generation of the FRB  emission are discussed in the literature, see e.g. \cite{2022ApJ...927....2K} and references therein.
Here we assume that the most feasible scenario is the generation of the maser emission at the reverse relativistic shock of a strong unmagnetized magnetar flare.

If a powerful flare hits a standing shock (which is assumed to be a termination shock (TS) of the wind), then a system of two relativistic shocks is to be formed. The forward shock (FS) propagates through the matter in the shocked wind medium, and the reverse shock (RS) -- through the material that forms the flare.  In the laboratory frame both shocks (and also the contact discontinuity -- CD hereafter) can move with a relativistic speed. To estimate these speeds one needs to consider the jump condition at each shock and the pressure balance at the CD.

Dynamics of the FS and the RS is discussed by \cite[][]{2022ApJ...927....2K}, we just adopt two key results from this paper
\citep[for a detailed discussion see][]{1976PhFl...19.1130B}. The bulk Lorentz factor of the shocks can be written as:
\be
\Gamma_{\rm fs}\approx\Gamma_{\rm rs}\approx \Gamma \approx \frac12\sqrt[4]{\frac{L_{\rm fl}}{L_{\rm sd}}}\,;
\label{eq:GGG}
\ee
and the flare penetration distance into the pulsar wind nebula is given by:
\be
\Delta R \approx \Delta t_{\rm fl}c\sqrt{\frac{L_{\rm fl} }{L_{\rm sd}}}\,.
\ee
Here $L_{\rm fl}$ is the flare luminosity and $L_{\rm sd}$ is the spin-down luminosity of the magnetar (note, that $L_{\rm sd}\approx L_\mathrm{pw}$). 
A typical energy of the radio emission associated with FRBs is \(\sim10^{40}\rm\,erg\). 
The maser mechanism can radiate away a per-cent fraction of  the kinetic energy of an ejecta of the magnetar flare \citep[see][and reference therein]{Koryagin2000}. 
It is natural to expected that the kinetic energy of the ejecta is larger than the energy emitted in X/gamma-rays. Thus, it is feasible that FRBs require magnetar flares of the total energy \(\sim10^{42}\rm\,erg\) and luminosity \(L_{\rm fl}\sim10^{44}\rm\,erg\,s^{-1}\) (given a 10~ms duration). This value is significantly smaller than the maximum recorded flare luminosity (see  Sec.~\ref{sec:intro}). 

Following \cite{2022ApJ...927....2K} one can obtain an equation for the typical frequency of the emission at the reverse shock: 
\be\label{eq:omaxrs}
\begin{split}
  \omega_{\rm max,rs}\approx & 2^{\nicefrac{-1}{8}}\frac{e}{m_ec^{\nicefrac{3}{2}}} \frac{L_{\rm fl}^{\nicefrac{3}{4}}}{L_{\rm sd}^{\nicefrac{1}{4}} R_{\rm TS}} \frac{(1-\sigma_{\rm fl})^{\nicefrac{3}{4}}}{ \Gamma_{\rm fl}\sigma_{\rm fl}^{\nicefrac{1}{4}}}\,\\
  \approx & 3\times10^{9}\,[{\rm rad \; s^{-1}}]\quad\times\\
  &L_{\rm sd,35}^{\nicefrac{-1}{4}} L_{\rm fl,45}^{\nicefrac{3}{4}} R_{\rm TS,15}^{-1}{ \Gamma_{\rm fl,3}^{-1}\sigma_{\rm fl,-2}^{\nicefrac{-1}{4}}}\,,
\end{split}
\ee
here \(\sigma_{\rm fl}\) is magnetization of the flare { and $R_\mathrm{TS}$ is the termination shock radius}.  
The expected position of the shock in the direction where  $\tau_9<1$, see eq.~(\ref{eq:Rs}) and also Sec.~\ref{sec:nmodel}, allows  to generate emission in the range from $\sim 100$ MHz up to  few GHz. 
 This emission mechanism produces a relatively narrow spectrum with a width $\Delta\nu/\nu \sim 0.2$.


\subsection{Evolution of the emitting site in the case of \frbs{}}
\label{sec:emdis}

In the paradigm of the maser cyclotron emission described in  Sec.~\ref{sec:Emission}, the emission frequency depends on the shock wave radius. In a massive binary system properties of the shock depend on  the orbital phase and the viewing angle $\phi$. The evolution of the shock wave radius is presented in  Fig.~\ref{fig:RTphidist}. As expected, the shock position  is  strongly anti-correlated with the optical depth which is presented in Fig.~\ref{fig:tauphitdist}.

The dependency of the typical frequency on the shock radius is given in eq.~(\ref{eq:omaxrs}). The shock radius changes during the orbital period, so does the frequency.
We illustrate the emission frequency evolution 
in Fig.~\ref{fig:nutdist} for $\dot{M}=4\times10^{-8}M_\odot$~yr$^{-1}$.  A slightly higher value of the mass loss is chosen to make the transparency window narrower to fit better parameters of the FRB 180916.J0158+65. 
Darker areas correspond to the orbital phases where $\tau_9>1$ while phases where $\tau_8>1$ are marked with a lighter colour.
White rectangles correspond to transparency ($\tau<1$) for the radio emission above 100 MHz. 
If flares have different values of the energy release and the Lorentz factor
then the emission frequency can vary significantly from flare to flare even at a given orbital phase. 
Independently on the eccentricity, inside a transparent window the frequency systematically drifts from high to low frequency ($\sim$1~GHz~$\rightarrow$~$\sim$100~MHz). This feature seems to be in correspondence with observations \citep{2021Natur.596..505P, 2021ApJ...911L...3P}.     


\subsection{Rotation measure evolution in the case of \frbl{}}
\label{sec:rme}

Rotation measure (RM) observations in the system FRB~121102 show $\mbox{RM}\approx 1.2\times10^5$~rad~m$^{-2}$ \citep[][]{2022MNRAS.511.6033P}. Moreover, the RM evolves on a time scale $t_{\rm RM}\sim 1$~yr.
The RM can be calculated as:
\be
 \mbox{RM} = \frac{e^3}{2\pi(m_ec^2)^2}\int_0^{R} n_e B \cos{\psi}dr.
\label{eq:RM0}
\ee
Here $\psi$ is the angle between the direction of a photon propagation and the magnetic field $B$, $n_e$ is the electron number density in the plasma.\footnote{Note, that the observed RM is given in [rad~m$^{-2}$]. Thus, it is necessary to use the normalization factor $10^4$ in eq.~(\ref{eq:RM0}) to obtain numbers in the same units.}   

In the framework of our model (independently on the exact emission mechanism), 
we can expect the following value of  RM in the stellar wind:
\be
\mbox{RM}_{\rm wind} \approx 140 \dot{M}_{-7.5} B_{-3} v_\mathrm{w,8.5}^{-1} \chi_{a,1.5}^{-1}  \mbox{ rad~m}^{-2} .
\label{eq:RMwind}
\ee
Here we adopt a magnetic field of the mG level. This value of the RM is close to a turbulent RM  necessary to explain depolarization of the signal at 1.4~GHz \citep[see ][]{2022MNRAS.511.6033P}. The maximum RM which we can obtain in the wind for the \frbl{} system is:
\be
\mbox{RM}_{\rm wind,max} \lesssim 8\times10^3 \dot{M}_{-7.5}^{3/2}\sigma_{\rm wind}^{1/2} v_\mathrm{w,8.5}^{-1/2} \chi_{a,1.5}^{-2}  \mbox{ rad \,m}^{-2} ,
\label{eq:RMwindmax}
\ee
where $\sigma_{\rm wind}$ is the magnetization of the stellar wind in the shocked region: $B_{\rm wind} = (8\pi\sigma_{\rm wind}p_{\rm wind})^{1/2}$. So, the observed value of RM in the \frbl{} is too high to be explained by a stellar wind.

The large value of observed RM can be explain by a propagation of the radio pulse through a raked shell of a supernova remnant (SNR) which expands into a dense interstellar medium (ISM) with $n\sim 100 $ cm$^{-3}$:
\be
\mbox{RM}_{\rm SNR} \approx 1.6\times 10^5 n_2^{11/10} E_{51}^{2/5} \sigma_{\rm SNR}^{1/2} t_{10.5}^{-1/5} \mbox{ rad~m}^{-2}. 
\label{eq:RMSNR}
\ee
Here the Sedov solution is applied and age is normalized to $10^{10.5}$~s (i.e., approximately 1 kyr).
In this model we assume amplification of the magnetic field up to the equipartition value \citep[in good agreement with observations, see ][]{2007Natur.449..576U}. Possible deviations from the equipartition can be accounted by the magnetization parameter $\sigma_{\rm SNR}$. 

The characteristic thickness of the shell at the Sedov state of a SNR evolution can be estimated as $\Delta R_{\rm SNR} \approx R_{\rm SNR}/20$. The variability time scale in the radial direction can be estimated as $t_{\rm RM,r}\approx \Delta R_{\rm SNR}/v_{\rm SNR,r}\sim 100\; t_{10.5}$~yr, where $t_{10.5}$ is the normalized SNR age. 
A  time scale of the RM evolution can be estimated from eq.~(\ref{eq:RMSNR}) as $t_{\rm RM,SNR} \approx 5000\;  t_{10.5}$~yr.  

There is a possibility to explain rapid changes of RM by a large scale turbulent motions related to the growth of the  Rayleigh-Taylor (RT)  instability and formation of RT fingers at the SNR contact discontinuity \citep[][]{2020A&A...642A..67T}. The appearance of one of the RT fingers on the line of sight can drastically change the RM on a relatively short time scale $\delta t \sim \delta r_{\rm RT finger}/c_\mathrm{s} \sim 1$~yr. Here we take $\delta r_{\rm RT finger} \sim 0.01  \Delta R_{\rm SNR} $ as the thickness of the transition zone between the RT finger and the SN ejecta. Also we assume $c_\mathrm{s} \sim v_{\rm SNR,r}$.
This behaviour with strong variations of RM on the time scale $\sim 1$~yr can repeat in future. 


\section{Discussion and conclusions}
\label{sec:dis}

Variations of the optical depth on the line of sight due to orbital motion modulate visibility of a repeating FRB flares for any magnetar-based emission model.
High orbital eccentricity allows to form very wide windows of transparency. Obviously,  they are wider for larger eccentricity (up to 80\% of the orbital period for $e=0.75$, see Fig.~\ref{fig:taudTT}). 

Large eccentricities are expected for systems with NSs (see, e.g. \citealt{2014LRR....17....3P} and references therein). 
It is natural to expect larger eccentricity for systems with longer orbital period. Thus, we can expect that width of the transparency window is larger, on average, in systems with long orbital periods. 
This can explain why in \frbl{} the width of the transparency window is larger \citep{2020MNRAS.495.3551R}.

{The drift from higher to lower frequency across the visibility window} \citep[see details in Fig.~9 of][]{2021ApJ...911L...3P} can be explained by a growth of the emitting region radius, which leads to a decrease of the frequency of the produced emission  (see eq.~\ref{eq:omaxrs} and Fig.~\ref{fig:nutdist}). 

Taking into account that a flare energy and its Lorentz factor can stochastically fluctuate from flare to flare, we can expect to obtain FRB properties similar to the observed in the case \frbs{}.
{In our model, the visibility window at higher frequency opens earlier and closes later. However, the typical frequency evolves due to changes in the position of the shock. This leads to relatively short duration of high frequency signal, despite the visibility window for high frequency radiation is still open. This explains why the fraction of the orbital period when a low frequency can be observed is larger.}

We suggest that the RM originated in the turbulent stellar wind can explain the frequency dependent polarization evolution in the case of \frbl{}.  Still, the stellar wind is unable to explain the  observed value RM$ = 1.2\times 10^5 \mbox{ rad~m}^{-2}$ \citep[][]{2022MNRAS.511.6033P}. 
Such huge value can be explained if the SN explosion happened close to a dense interstellar cloud with $n\sim 100$~cm$^{-3}$.
Note, that a cloud can also absorb optical emission preventing direct detection of the normal component of the binary system.

In the framework of the maser cyclotron radiation mechanism  an isolated magnetar might produce FRB radio emission roughly at the same distance for different flares. According to eq.~(\ref{eq:omaxrs}) the typical frequency will not change {from flare to flare  as significantly as in binary systems with strong winds}. As the spectrum width is characterized by $\Delta \nu/ \nu \sim 0.2$, we can expect that radio bursts would not be simultaneously detected at significantly different frequencies, e.g. $\sim100-200$~MHz and $\sim 1$~GHz (which corresponds, e.g. to LOFAR and Westerbork/Apertif, see \citealt{2021Natur.596..505P} about simultaneous observations with these instruments). 
 
In this study, we assume a spherically symmetric stellar wind and so, do not discuss topics related to the stellar wind asymmetry. If the normal component of a binary is a Be star then the picture of the winds interaction can be much more complicated in comparison to the one discussed above. This requires a large scale 3D R(M)HD modeling.  
Still, we conclude that main properties of FRBs which demonstrate long-term periodicity can be described in the framework of a magnetar in a massive binary system, where a spiral-like structure formed due to stellar and pulsar winds interaction explains a duty cycle via formation of windows of transparency at particular orbital phases.


\section*{Acknowledgements}

The calculations were carried out in the CFCA cluster XC50 of National Astronomical Observatory of Japan. 
Authors are grateful to D.~Khangulyan for his contribution at the initial phase of this project.
We also thank the referee for useful comments.
SBP acknowledges support from the Russian Science Foundation, grant 21-12-00141 and  BMV acknowledges NASA grant 80NSSC20K1534.

\section*{Data Availability}

 Observational data used in this paper are quoted from the
cited works. Data generated from computations are reported
in the body of the paper. Additional data can be made available upon reasonable request.




\bibliographystyle{mnras}
\bibliography{frb} 








\bsp	
\label{lastpage}
\end{document}